%% file: laplace_transport_map_arxiv.tex
\pdfoutput=1
\documentclass[11pt]{article}

\usepackage[margin=1in]{geometry}
\usepackage{setspace}
\usepackage{amsmath,amssymb,bm}
\usepackage{graphicx}
\usepackage[authoryear,round]{natbib}
\usepackage{authblk}
\usepackage{url}
\usepackage[hidelinks]{hyperref}
\usepackage{algorithmic, algorithm}

\newcommand{\introduction}{\section{Introduction}}
\newcommand{\conclusions}{\section{Conclusions}}
\newcommand{\codedataavailability}[1]{\section*{Code and data availability}#1}
\newcommand{\authorcontribution}[1]{\section*{Author contributions}#1}
\newcommand{\competinginterests}[1]{\section*{Competing interests}#1}
\newcommand{\financialsupport}[1]{\section*{Financial support}#1}
\newenvironment{acknowledgements}{\section*{Acknowledgements}}{}

\usepackage{bm}
\renewcommand{\vec}[1]{\bm{#1}}

\title{Laplace Bayesian transport maps for generative modeling of non-Gaussian climate fields}

\author[1]{Jacob A. Johnson}
\author[1]{Felix Jimenez}
\author[1]{Matthias Katzfuss\thanks{Corresponding author: katzfuss@gmail.com}}
\affil[1]{Department of Statistics, University of Wisconsin--Madison, Madison, WI, USA}
\date{}

\begin{document}
	
	\maketitle
	
	\input{00_abstract}
	\input{01_intro}
	\input{02_background}
	\input{03_methodology}
	\input{04_numerical_comparison}
	\input{05_conclusion}

	\codedataavailability{The transport maps were fitted with the Python package batram (\url{https://github.com/katzfuss-group/batram}), and the exact version used
		is included in the code archive.
		Code to reproduce all analyses and figures is publicly available at \url{https://doi.org/10.5281/zenodo.24203962} \citep{Johnson2026code}. 
		The precipitation fields from Sect.~\ref{sec:application_data} were taken from the Community Earth System Model (CESM) Large Ensemble Project, available at \url{https://doi.org/10.5065/D6J101D1} \citep{ncar_gdex_dataset_d651027}. The
		extracted 1 July fields used in this study are included in the code archive as well.}
	
	\authorcontribution{JJ implemented the methods, carried out the numerical experiments, and wrote the first draft of the manuscript. 
		FJ helped with an initial implementation of the numerical experiments and provided JJ with guidance on the prior methodology and software. 
		MK initiated and supervised the project and revised the manuscript.}
	
	\competinginterests{The authors declare that they have no conflict of interest.}
	
	\begin{acknowledgements}
		We would like to thank Paul Wiemann for helpful comments and discussions.
		The numerical experiments used the computing resources of the Center for High Throughput Computing at the University of Wisconsin--Madison \citep{CHTC2006}.
		In doing this work, the authors used Claude (Opus 5.5; Anthropic) and Gemini (Thinking 3.6; Google) as research-assistance tools. 
		These tools assisted with writing and debugging code for implementing the methods, post-processing results, and producing figures. All code was reviewed, tested, and run by the authors, and all numerical results and figures were produced from the authors’ own computations.
		The tools were also used to explore alternative methodological approaches and to improve the clarity and organization of the manuscript. 
		All references were obtained and checked from the original sources. 
		The manuscript, methodological decisions, and the interpretation of the results were reviewed, verified, and approved by the authors, who take full responsibility for the content of this article.
		
	\end{acknowledgements}
	
	\financialsupport{The authors were partially supported by NASA's Advanced Information Systems Technology Program (AIST-21 and AIST-23, the latter under grant 80NSSC26K0595), by National Science Foundation (NSF) Grant DMS-1953005/2433548, and by industry members of the Center for Interdisciplinary Research on Convective Storms (CIRCS) and NSF award numbers 2517152 and 2517615. MK's research was also partially supported by the Office of the Vice Chancellor for Research at the University of Wisconsin--Madison with funding from the Wisconsin Alumni Research Foundation.}
	
	\bibliographystyle{copernicus}
	\bibliography{mendeley,additionalrefs}
	
\end{document}

%% file: 00_abstract.tex
\begin{abstract}
Transport maps provide a powerful framework for the generative modeling of complex, non-Gaussian spatial fields, such as those encountered in climate science. By mapping a complicated target distribution to a standard Gaussian reference distribution, they can serve as efficient stochastic surrogates. However, current Bayesian transport maps, which are modeled using autoregressive Gaussian processes, typically rely on empirical Bayes estimates that ignore hyperparameter uncertainty. Here, we contrast this point-estimate approach with a fully Bayesian model that provides posterior uncertainty quantification in transport map hyperparameters. To avoid the high cost of Markov chain Monte Carlo (MCMC), we also propose a Laplace approximation to the hyperparameter posteriors, which can be computed at negligible additional computational cost relative to empirical Bayes. We provide numerical evidence and a real-data climate application demonstrating that these Laplace approximations can accurately capture most posterior summaries at a fraction of the cost of MCMC, while maintaining good predictive properties. Thus, our approach offers a scalable, uncertainty-aware framework for the generative modeling of non-Gaussian climate fields.
\end{abstract}

%% file: 01_intro.tex
\introduction\label{sec:intro}

\begin{figure}
    \centering
    \includegraphics[width =\linewidth]{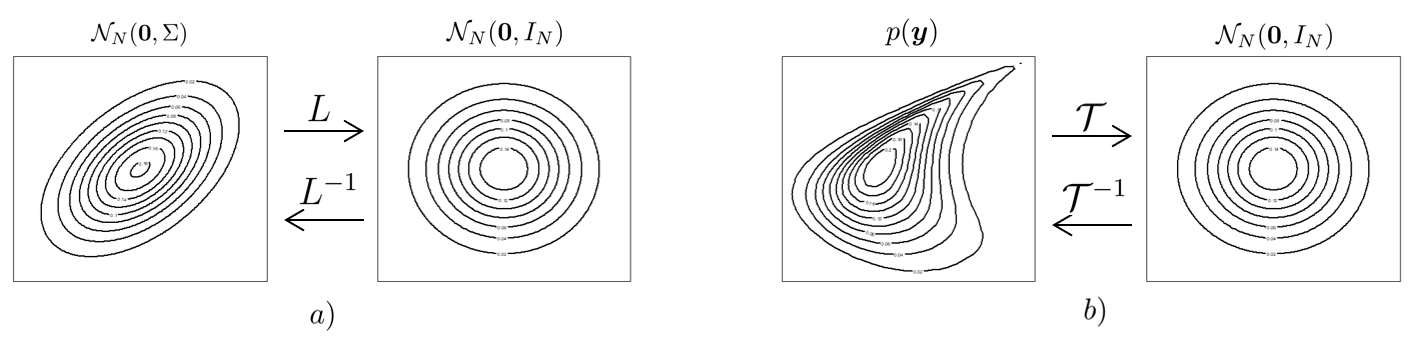}
    \caption{Similar to how a triangular inverse Cholesky factor $\mathbf{L}$ maps a non-standard Gaussian random variable to a standard Gaussian (a), a triangular transport map can transform a non-Gaussian distribution to a standard-Gaussian reference distribution (b). Because the map is invertible, one can convert the standard Gaussian back to the target distribution of interest (e.g., for sampling).}
    \label{fig:map_example}	
\end{figure}

Modeling non-Gaussian spatial fields is important in many scientific applications, such as building stochastic surrogates for climate models \citep[e.g.,][]{Castruccio2014,Nychka2018} 
and generative modeling of environmental variables such as groundwater \citep[e.g.,][]{HORNING2025100872}, ozone \citep[e.g.,][]{10.1093/jrsssc/qlac007}, vegetation \citep[e.g.,][]{10.1093/forestscience/57.3.212}, and wind speed \citep[e.g.,][]{https://doi.org/10.1002/env.2632}. Transport maps have emerged as a powerful tool for generative modeling, capable of transforming these complicated distributions into simpler, more tractable forms \citep[e.g.,][]{Marzouk2016}. Similar to how a triangular inverse Cholesky factor maps a Gaussian distribution to a standard Gaussian, a triangular transport map is able to map a non-Gaussian multivariate distribution to a simple reference distribution, usually Gaussian (see Fig.~\ref{fig:map_example}). These triangular maps are a type of autoregressive normalizing flow \citep{Papamakarios2021} that relies on Gaussian processes rather than the highly parameterized neural networks typically used. An invertible transport map serves as a stochastic surrogate by mapping the reference distribution back to the original non-Gaussian space. This generative capability makes sampling and other operations possible by exploiting the well-behaved properties of the reference distribution.

Previous works using transport maps to model spatial fields \citep[e.g.,][]{Katzfuss02042024} have used an empirical Bayes approach to optimize a finite set of Gaussian process hyperparameters to fit the transport map to training data. Once the map is estimated using the training data, it is then assumed to be fixed and known when used in further inference. In this case, the transport map can be learned using gradient-based optimization methods and automatic differentiation. As discussed in \citet{Katzfuss02042024}, this approach has been favored over a fully Bayesian approach that finds the posterior distribution of hyperparameters via Markov chain Monte Carlo (MCMC) due to the computational cost of sampling and the inconvenience of having no closed-form expression for the map.
While this does make the task of using transport maps less computationally demanding, it does have some significant drawbacks. 

Assuming that the transport map is fixed, when it was in fact estimated from training data, may lead to suboptimal model performance, overconfident prediction intervals, and/or bias when generating samples from the target distribution. However, allowing for uncertainty on the transport map structure is difficult because many currently available results have only been proven for a fixed transport map, and inference on the uncertain map may be difficult because results may not be available in closed form. In the Bayesian paradigm, when results are not available in closed form, practitioners often turn to MCMC techniques \citep[see][for an introduction]{speagle2020conceptualintroductionmarkovchain}. The computational burden of fitting models to large spatial fields is already well documented \citep[e.g.,][]{Heaton2017}, and MCMC methods only compound that computational load by requiring repeated evaluations of the model likelihood. 

Our goal in this work is to explore the uncertainty that comes from fitting the map with training data while evaluating the use of various uncertainty quantification methods across various sample sizes and spatial domains. We propose a Laplace approximation \citep[see][for an introduction]{Amaral_Turkman_Paulino_Müller_2019} to the posterior distribution of the estimated map hyperparameters, which is computationally efficient while still describing the uncertainty in map behavior. Other approaches such as \citet{chakraborty2024learningnongaussianspatialdistributions} have focused on using shrinkage priors to shrink the map toward a ``base'' parametric Gaussian family, but they still assume that the map is fixed after shrinkage. We know of no other work that examines the uncertainty quantification of the transport map itself in the setting of modeling spatial fields.

Our contribution is to build Bayesian transport maps by incorporating a Laplace approximation to the posterior distribution of the hyperparameters that define the map, allowing us to exploit the benefits of the Bayesian framework in a computationally efficient way. This will enhance the flexibility and robustness of our generative model for non-Gaussian spatial fields, even with relatively few training samples. We demonstrate the effectiveness and drawbacks of this approximation and discuss its effect on the transport map components. We specifically highlight the utility of this approach for climate science through a real-data application involving global precipitation rates.

The remainder of this document is organized as follows. In Sect.~\ref{sec:background}, we review the notation of a Bayesian transport map, focusing on how such a map can be characterized by a small set of hyperparameters. Next, in Sect.~\ref{sec:methodology}, we detail the methods of applying uncertainty quantification to the transport map model. Then, in Sect.~\ref{sec:numerical}, we compare the fixed transport map to the uncertainty-aware maps in a simulation study and in a real-data climate application. Last, in Sect.~\ref{sec:conclusion}, we conclude and discuss our findings.

%% file: 02_background.tex
\section{Background}\label{sec:background}

\subsection{Transport maps}
Let $\vec{y} = (y_1, \dots, y_N)^\top$ denote an $N$-dimensional random vector (spatial field) with joint (non-Gaussian) distribution $p(\vec{y})$, which we assume to be centered to have mean $\vec{0}$.
There exists a measurable bijection, a lower-triangular transport map, $\mathcal{T}: \mathbb{R}^N \to \mathbb{R}^N$, that maps $p(\vec{y})$ to a simpler reference distribution \citep{Marzouk2016}, for which we will use a standard multivariate normal:
\begin{equation} \label{map}
    \vec{z} = \mathcal{T}(\vec{y}) \sim \mathcal{N}_{N}(\vec{0}, \mathbf{I}_N) \quad \text{if} \quad \vec{y} \sim p(\vec{y}),
\end{equation}
\begin{equation}
    \mathcal{T}(\vec{y}) = \begin{pmatrix}
\mathcal{T}_1(y_1) \\
\mathcal{T}_2(y_1, y_2) \\
\vdots \\
\mathcal{T}_N(y_1, \dots, y_N)
\end{pmatrix}.
\end{equation}

The transport map $\mathcal{T}$ is referred to as a triangular
transport map if the $i$th component of the transport map is only a function of $y_1,\ldots,y_i$, according to some ordering. Section~\ref{subsec::ordering} describes the ordering we recommend. The triangular structure ensures that the Jacobian of the transformation $\nabla\mathcal{T}(\vec{y})$ is lower-triangular \citep[e.g., see][]{doi:10.1137/080740647}, meaning that the density can be calculated relatively efficiently. 

\subsection{Factorization and map parameterization}
The triangular nature of the map $\mathcal{T}$ and its Jacobian allows the distribution of $\vec{y}$ to be easily factored into a series of conditional distributions. Using the change-of-variables formula, each conditional can be expressed in terms of the reference (standard Gaussian) density and the derivative of its associated transport map component: 
\begin{equation*}
    p(\vec{y}) = \prod^N_{i=1}p(y_i|\vec{y}_{1:i-1}) = \prod^N_{i=1}\left(p_z(\mathcal{T}_i(\vec{y}_{1:i})) \left\vert \frac{\partial \mathcal{T}_i(\vec{y}_{1:i})}{\partial y_i} \right\vert \right).
\end{equation*}
Thus, we have turned the evaluation of our non-Gaussian target density into a product of univariate standard normal densities, $p_z(\cdot)$, along with the associated partial derivatives that complete the transformation. This factorization facilitates efficient sequential generation of the field, as each location can be sampled conditionally on the previous ones.

Following \citet{Katzfuss02042024}, we parameterize each component of the transport map as 
\begin{equation*}
\mathcal{T}_i(\vec{y}_{1:i}) = \frac{y_i - f_i(\vec{y}_{1:i-1})}{d_i},
\end{equation*}
where $f_i: \mathbb{R}^{i-1} \to \mathbb{R}$ is a regression function, and $d_i > 0$ is a conditional standard deviation for each $y_i$. This allows the joint density of $\vec{y}$, conditional on $\vec{f} = (f_1,\ldots,f_N)$ and $\vec{d} = (d_1,\ldots,d_N)$, to be expressed as
\begin{equation*}
p(\vec{y} | \vec{f},\vec{d}) =\prod^N_{i=1}p(y_i|\vec{y}_{1:i-1},f_i,d_i)= \prod_{i=1}^N \mathcal{N}\left(y_i \mid f_i(\vec{y}_{1:i-1}), d_i^2\right).
\end{equation*}

\subsection{Ordering, sparsity, and scalability}\label{subsec::ordering}
Using the notation and structure outlined above, the transport map has an autoregressive structure, where each successive map component only relies on previously ordered data points. However, this can be computationally expensive, because the regression function $f_i$ involves all $i-1$ of the previous $y_j$'s.
We can mitigate some of these computational issues by assuming conditional independence in $p(\vec{y})$:
\begin{equation}
    \mathcal{N}\left(y_i \mid f_i(\vec{y}_{1:i-1}), d_i^2\right) \approx \mathcal{N}\left(y_i \mid f_i(\vec{y}_{c(i)}), d_i^2\right), \; c(i) \subset \{1,\dots,i-1\},\ |c(i)| = \min(i-1,m). \label{eq:sparsity}
\end{equation}
Here, $c(i)$ is a small conditioning set of at most size $m$. This conditional independence enforces sparsity in the Jacobian of the map $\mathcal{T}$. 
This process parallels Vecchia approximations of Gaussian processes, which are common in spatial statistics \citep[see][]{Katzfuss&Guinness2021}. A sparse triangular map is therefore a nonlinear, non-Gaussian generalization of a sparse inverse Cholesky factor. 

A critical modeling choice in constructing the sparse triangular transport map is the ordering of the data. This will affect both the structure of the map in Eq.~\eqref{map} and how sparsity is introduced through the conditioning sets described in Eq.~\eqref{eq:sparsity}. Because our focus is on modeling spatial fields, we will focus on creating an ordering based on the spatial location of the data. \citet{Guinness2016a} describes various ordering schemes, of which we will use the \emph{maximum-minimum-distance} (maximin) ordering, where locations are ordered by maximizing the minimum distance to all previously ordered locations. First, we choose an initial location, indexed by $i_1$. Then, if $\vec{s}_{i_j}$ is the spatial location of observation $y_{i_j}$, we choose the next location, $\vec{s}_{i_{j+1}}$, by $i_{j+1} = {\arg \max}_{k \notin \{i_1,\ldots,i_j\}} \min_{r \leq j} \| \vec{s}_k - \vec{s}_{i_r} \|$. 

After reordering the locations according to the maximin ordering, we let
$r_i = \min_{k < i} \lVert \vec{s}_i - \vec{s}_k \rVert$, $i = 2,\ldots,N$, denote the distance from location $i$ to its nearest previously ordered location, which is nonincreasing in $i$ by construction of the ordering. Because $r_1$ is undefined, we set
$r_1 = r_2^2 / r_6$ and work with the normalized distances $\ell_i = r_i / r_1$ for $i = 1,\ldots,N$, so that $\ell_1 = 1$ and $1 \geq \ell_2 \geq \cdots \geq \ell_N > 0$.

\begin{figure}
    \centering
    \includegraphics[width =\linewidth]{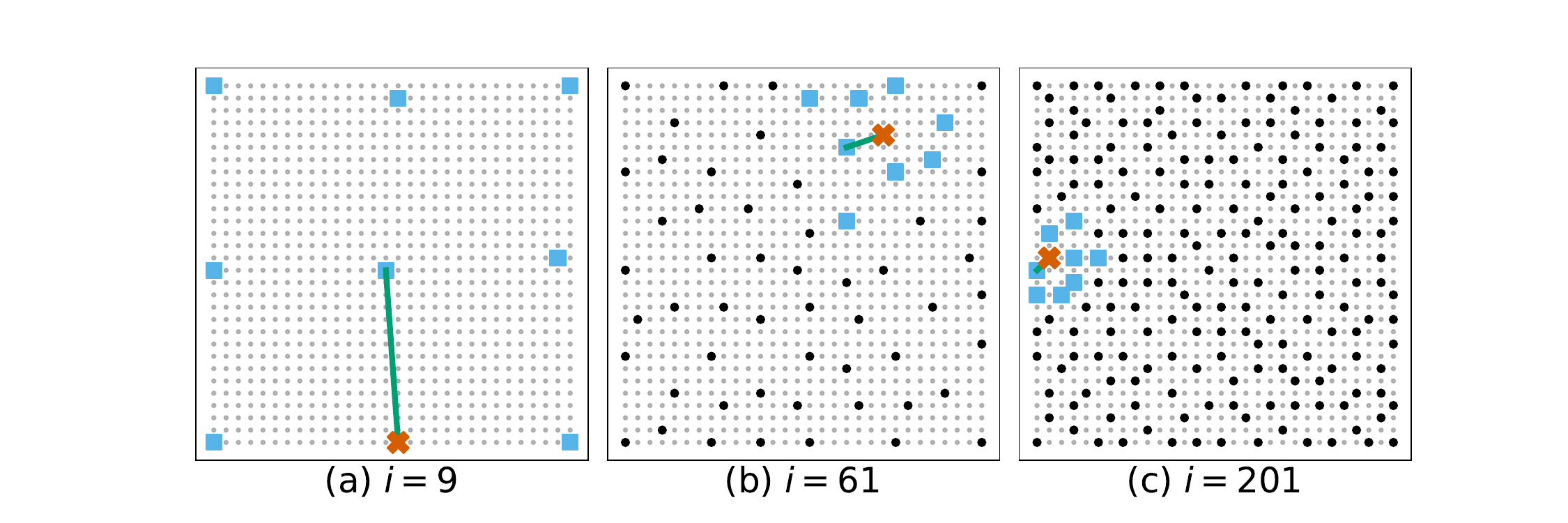}
    \caption{Maximin ordering for locations on a $30 \times 30$ simulated grid on the unit square. Small gray points show all locations, the black points show the previously ordered locations, the orange X shows the $i$th location, and the blue squares show the closest $m=8$ neighbors. The green lines show $r_i$, the distance to the closest previously ordered location.}
    \label{fig:max-min}	
\end{figure}

By defining our ordering in this way (see Fig.~\ref{fig:max-min}), the ordered data spread early points across the domain, ensuring that subsequent conditioning sets are both diverse and representative of the field as a whole. Given this ordering, the conditioning set $c(i)$ for each location is defined as the $m$ nearest previously ordered points, using the same distance that was used to compute the ordering. Figure~\ref{fig:max-min} illustrates the construction of the ordering and conditioning sets using $m=8$. The choice of $m$ is a balance between accuracy and scalability; larger $m$ allow for a better approximation, while smaller $m$ yield sparser maps with their reduced computational cost. One could choose to fix $m$, treating it as a tuning parameter that is chosen based on available computational resources and/or desired approximation quality. However, we choose to define $m$ as a function of a hyperparameter that we will learn from training data (see Sect.~\ref{sec:priors} for details), allowing the training fields to influence the number of previously ordered neighboring points in the conditioning set.

\subsection{Priors and map hyperparameters}
\label{sec:priors}
To flexibly model the regression functions, we place independent Gaussian-process inverse-Gamma priors on each pair $\left(f_i,d_i^2\right)$:
\begin{align*}
d_i^2 &\sim \text{Inverse-Gamma}(\alpha_i, \beta_i)\\
f_i \mid d_i &\sim \mathcal{GP}\left(0, d_i^2 K_i(\cdot, \cdot)\right)
\end{align*}
where $K_i(\cdot, \cdot)$ is a chosen covariance/kernel function (e.g., Mat\'ern). 
The full set of prior parameters can then be written as $\vec{\Psi} = \begin{pmatrix}
\alpha_1, \dots, \alpha_N, \beta_1, \dots, \beta_N, \vec{\kappa}_1, \dots, \vec{\kappa}_N
\end{pmatrix}$, where $\vec{\kappa}_i$ is the vector of parameters that determine $K_i(\cdot, \cdot)$. 

Because the size of $\vec{\Psi}$ grows with $N$, we make some additional assumptions in order to express $\vec{\Psi}$ as a function of a smaller parameter vector $\vec{\theta}$ of constant size. In the setting of spatial fields, \citet{Schafer2017} and \citet{Katzfuss02042024} detail a power-functional decay of the variance $\operatorname{Var}(y_i \mid f_i, d_i, \vec{y}_{1:i-1}) = d_i^2$ with respect to $\ell_i$, the normalized distance to the nearest neighbor to point $i$. So, the prior on $d^2_i$ can be specified to reflect this decreasing variance by setting
\begin{equation}
    \alpha_i = 2 + g^{-2}, \qquad
    \beta_i = \left(1 + g^{-2}\right) \exp\left(\theta_{d_1} + \theta_{d_2} \log \ell_i\right),
    \label{eq:prior_di}
\end{equation}
which gives $\operatorname{E}(d_i^2) = \beta_i/(\alpha_i - 1) = \exp(\theta_{d_1} + \theta_{d_2}\log\ell_i)$
and $\operatorname{SD}(d_i^2) = \operatorname{E}(d_i^2)/\sqrt{\alpha_i - 2} = g\,\operatorname{E}(d_i^2)$. In this work, we set $g=4$, which results in a relatively uninformative prior on each $d_i^2$ that depends on only two hyperparameters, $\theta_{d_1}$ and $\theta_{d_2}$.

Additionally, we make some assumptions on the form of the covariance function $K_i(\cdot, \cdot)$. We assume that $K_i(\cdot, \cdot) = C_i(\cdot, \cdot) / \operatorname{E}(d^2_i)$ where 
\begin{equation}
    C_i(\vec{y}_{1:i-1}, \vec{y}_{1:i-1}^{'}) = \vec{y}_{1:i-1}^{\top} \mathbf{Q}_i \vec{y}_{1:i-1}^{'} + \sigma_i^2 \rho_i\left(\frac{\sqrt{\left(\vec{y}_{1:i-1} - \vec{y}_{1:i-1}^{'}\right)^{\top}\mathbf{Q}_i\left(\vec{y}_{1:i-1} - \vec{y}_{1:i-1}^{'}\right)}}{\gamma}\right), \hspace{1em} i = 1, \dots, N.
    \label{eq:kernel}
\end{equation}
Here, $\rho_i$ is a positive definite correlation function, which we take to be the Mat\'ern correlation function with smoothness $\nu=1.5$ for all $i$. The range parameter is $\gamma = \exp(\theta_{\gamma})$, and we assume that $\sigma_i$ decays similarly to $d_i$, so we parameterize $\sigma_i = \exp(\theta_{\sigma_1} + \theta_{\sigma_2}\log(\ell_i))$. 
Note that the value of $\sigma_i$ controls the nonlinearity of the regression function $f_i$. This parameterization allows the degree of nonlinearity to decrease along with $\ell_i$ (whenever $\theta_{\sigma_2} > 0$), in which case the regression functions become more linear and the conditional distributions of $y_i | \vec{y}_{1:i-1}$ become more Gaussian as we get further in the ordering.

Lastly, we construct the $(i-1) \times (i-1)$ diagonal matrix $\mathbf{Q}_i$. With the entries of $\vec{y}_{1:i-1}$ in Eq.~\eqref{eq:kernel} arranged by increasing distance to location $i$, we define $q_j = \exp\bigl(-j \exp(\theta_q)\bigr)$ as a relevance weight for the $j$th nearest previously ordered location. This reflects the prior belief that the relevance of previously ordered points will decay as they get further away. As mentioned in Sect.~\ref{subsec::ordering}, we assume that the regression functions can be approximated by a function that only relies on a small subset of the previously ordered points. We achieve this by inducing additional sparsity into $\mathbf{Q}_i$ by setting 
\begin{equation}
    (\mathbf{Q}_i)_{j,j} = \begin{cases} q_j, & j \leq m,\\
    0, & j > m.     \end{cases}
\end{equation}
Because the diagonal entries of $\mathbf{Q}_i$ are only non-zero for neighbors in $c(i)$, both terms of Eq.~\eqref{eq:kernel} depend on
$\vec{y}_{1:i-1}$ only through $\vec{y}_{c(i)}$, so $f_i$ only depends on the nearest previously ordered points contained in $c(i)$. This yields the conditional-independence structure in Eq.~\eqref{eq:sparsity}. 

We determine the number of previously ordered neighbors $m$, as used here and in Sect.~\ref{subsec::ordering}, from $\theta_q$ by choosing $m = \max\left\{j : q_j^{1/2} \geq \epsilon\right\} =\left\lfloor -2 \log(\epsilon)\, e^{-\theta_q} \right\rfloor$ for some $\epsilon > 0$, meaning that we are only relying on the previously ordered points whose relevance, as measured by the square root of the corresponding diagonal entry of $\mathbf{Q}_i$, is at least $\epsilon$. In all our numerical examples, we take $\epsilon = 0.01$, meaning that $m \geq 1$ requires $\theta_q \leq \log(-2\log 0.01) \approx 2.22$.

In summary, using the carefully constructed priors detailed above, we have reduced our hyperparameter space from $\vec{\Psi}$, which grew with $N$, to the small set of hyperparameters $\vec{\theta} = \begin{pmatrix}
\theta_{d_1},
\theta_{d_2},
\theta_{\sigma_1}, 
\theta_{\sigma_2},
\theta_{\gamma},
\theta_{q}
\end{pmatrix}^{\top}$ of fixed size. For further details about the motivations for these prior assumptions in the case of spatial fields, we refer to \citet{Katzfuss02042024,Wiemann2023}. 

Our goal is to find the posterior distribution of the parameters $\vec{\theta}$ given $n$ independent training samples $\vec{y}^{(1)},\dots, \vec{y}^{(n)}$, which we organize into an $n\times N$ matrix $\mathbf{Y}$. Given $\vec{\theta}$, the pairs $(f_i, d_i)$
can be integrated out, which yields the \emph{integrated likelihood}
\begin{equation}
	p(\mathbf{Y} \mid \vec{\theta})
	= \prod_{i=1}^N \int \prod_{t=1}^n
	p\bigl(y_i^{(t)} \mid \vec{y}_{c(i)}^{(t)}, d_i, f_i\bigr)\,
	p(f_i, d_i \mid \vec{\theta})\, \mathrm{d}d_i\, \mathrm{d}f_i .
	\label{eq:likelihood}
\end{equation}

Because the prior on $(f_i, d_i)$ is conjugate, each integral in Eq.~\eqref{eq:likelihood} is available in closed form, making the $i$th factor a multivariate $t$ density for the $n$ training values at location $i$, whose scale matrix depends on the kernel $K_i$ evaluated on the corresponding conditioning vectors \citep[][Proposition~2]{Katzfuss02042024}. This closed form provides substantial computational benefits \citep[see][for implementation details]{Wiemann2023}.

To complete the model, we place independent Gaussian priors $\theta_k \sim \mathcal{N}(\mu_k, \tau^2)$ on each hyperparameter. A proper prior is needed because the integrated likelihood in Eq.~\eqref{eq:likelihood} does not decay to zero in every direction of the parameter space. For example, as $\theta_{\sigma_1} \to -\infty$, $\sigma_i \to 0$ for all $i$, so the nonlinear term in Eq.~\eqref{eq:kernel} vanishes and the regression functions $f_i$ become linear. Therefore, as $\theta_{\sigma_1}$ decreases, $p(\mathbf{Y} \mid \vec{\theta})$ converges to a strictly positive limit that does not depend on $\theta_{\sigma_2}$ or $\theta_\gamma$.
Once each $\sigma_i$ is negligible, changing $\theta_{\sigma_1}$ further no longer changes the model, so an improper uniform prior would assign posterior density bounded away from zero to an infinitely long region of the parameter space, leading to an improper posterior. This mirrors well-known results for uniform priors on the range parameters of Gaussian processes \citep{Berger2001_new} and on variance components in linear mixed models \citep{Hobert01121996}. While the problem is most pronounced for small $n$, it does arise for all $n$, so we use diffuse normal priors instead of an improper uniform prior. 

We set the prior mean as follows. First, $\mu_{d_1}=\log \bar{v}_1$,  where $\bar{v}_1$ is the average squared (centered) training value at the first location in the ordering. This corresponds to a prior map in which $\operatorname{E}(d_1^2) = \bar{v}_1$, chosen empirically. Next, $\mu_{d_2} = 0.2$, meaning that the conditional variances decay weakly with the ordering. With $\mu_{\sigma_1} = \mu_{\sigma_2} = 0$ we specify that the degree of nonlinearity is constant along the ordering and that $\sigma_i = 1$. The prior mean $\mu_{\gamma}=-1$ means that the range parameter is $\gamma = e^{-1}$, and lastly, $\mu_{q}=0$ centers the chosen number of neighbors at $m = 9$. Along with specifying the means, we set $\tau = 10$, which is very diffuse given that all elements of $\vec{\theta}$ are on a log scale or are exponents
of $\ell_i$. 

Combining the integrated likelihood in Eq.~\eqref{eq:likelihood} with the prior, the posterior distribution of the hyperparameters is $p(\vec{\theta} \mid \mathbf{Y}) \propto p(\mathbf{Y} \mid \vec{\theta})\, p(\vec{\theta})$.
Although each factor of $p(\mathbf{Y} \mid \vec{\theta})$ is available in closed form, the posterior itself is not, because $\vec{\theta}$ enters the likelihood nonlinearly through $\beta_i$ and $K_i$. The various methods that we employ for working with and approximating this posterior are detailed in Sect.~\ref{sec:methodology}.

%% file: 03_methodology.tex
\section{Methodology}\label{sec:methodology}

\subsection{Empirical Bayes approach}\label{emp_bayes}

Most current work on Bayesian transport maps for spatial fields such as \citet{chakraborty2024learningnongaussianspatialdistributions,Katzfuss02042024,Wiemann2023} take an empirical Bayes approach to estimating the hyperparameters in $\vec{\theta}$. Hyperparameters are selected via optimization and then fixed while further inference is conducted. This method is convenient because many results have been proved for a fixed transport map, and the initial optimization can be done using stochastic gradient methods, which are simple and efficient thanks to tools like automatic differentiation (autodiff) \citep[see][for details of an application in PyTorch]{Paszke2017AutomaticDI}. 

Standard empirical Bayes approaches \citep[e.g.,][]{Katzfuss02042024} select the parameters $\hat{\vec{\theta}}$ using gradient-based minimization of the negative log integrated likelihood 
\begin{equation}
    \mathcal{L}(\vec{\theta}) = -\log p(\mathbf{Y} \vert \vec{\theta}) = \sum_{i=1}^{N} \mathcal{L}_i(\vec{\theta})
    \approx \frac{N}{|\mathcal{B}|} \sum_{i \in \mathcal{B}} \mathcal{L}_i(\vec{\theta}) = \hat{\mathcal{L}}_{\mathcal{B}}(\vec{\theta})
    \label{eq:intlik}
\end{equation}
as a function of $\vec{\theta}$, where $\mathcal{L}_i(\vec{\theta})$ is the negative log integrated likelihood of the $i$th map component, i.e., of location $i$ in the maximin ordering. The location-wise decomposition of Eq.~\eqref{eq:intlik} allows for minibatching over a batch $\mathcal{B} \subset \{1, \dots, N\}$ of locations instead of the full $ \mathcal{L}(\vec{\theta})$ if desired.

In this work, rather than optimizing only the negative log integrated likelihood, we choose to use the full negative log-posterior which includes the prior on $\vec{\theta}$ described in Sect.~\ref{sec:priors}. The objective function can be written as the negative log posterior, up to an additive constant,
\begin{equation}
    \tilde{\mathcal{L}}(\vec{\theta}) = \mathcal{L}(\vec{\theta}) - \log p(\vec{\theta}) = \left( \sum_{i=1}^{N} \mathcal{L}_i(\vec{\theta}) \right) - \log p(\vec{\theta}) 
    \approx \left( \frac{N}{|\mathcal{B}|} \sum_{i \in \mathcal{B}} \mathcal{L}_i(\vec{\theta}) \right) - \log p(\vec{\theta})  = \hat{\tilde{\mathcal{L}}}_{\mathcal{B}}(\vec{\theta}) ,
    \label{eq:L_tilde}
\end{equation}
which is used to find the optimal value 
\begin{equation}
    \hat{\vec{\theta}} = \underset{\vec{\theta}}{\operatorname{argmax}} \; p(\vec{\theta} \mid \mathbf{Y})  = \underset{\vec{\theta}}{\operatorname{argmin}} \, \tilde{\mathcal{L}}(\vec{\theta}).
    \label{eq:theta_hat}
\end{equation}
Because our chosen prior on $\vec{\theta}$ is very diffuse, the estimates of $\vec{\theta}$ based on $\tilde{\mathcal{L}}(\vec{\theta})$ and on $\mathcal{L}(\vec{\theta})$ are very close once the number of training fields becomes large enough ($n\geq 15$ in our numerical experiments). 

Once $\hat{\vec{\theta}}$ is found, the posterior $p(\vec{\theta} \mid \mathbf{Y})$ is fixed to be a point mass at $\hat{\vec{\theta}}$, so $p(\vec{\theta} \mid \mathbf{Y}) = \delta_{\hat{\vec{\theta}}}$. This point mass distribution has the same effect as assuming that the map is fixed with $\vec{\theta} = \hat{\vec{\theta}}$. The algorithm for fitting the transport map using empirical Bayes is included below as Algorithm~\ref{alg:empbayes}. 

\begin{algorithm}[htb]
\caption{Empirical Bayes model fitting}
\label{alg:empbayes}
\begin{algorithmic}[1]
\REQUIRE Training data $\mathbf{Y}$, prior $p(\vec{\theta})$, learning rate $\eta$, number of epochs $E$
\ENSURE $\hat{\vec{\theta}}$
\STATE Initialize hyperparameters $\vec{\theta}$
\FOR{$e=1,2,\ldots,E$}
 \STATE Compute $\nabla_{\vec{\theta}} \tilde{\mathcal{L}}(\vec{\theta})$ using autodiff, where $\tilde{\mathcal{L}}(\vec{\theta})$ is defined in Eq.~\eqref{eq:L_tilde}
 \STATE $\vec{\theta} \gets \vec{\theta} - \eta \,\nabla_{\vec{\theta}} \tilde{\mathcal{L}}(\vec{\theta})$
\ENDFOR
\STATE $\hat{\vec{\theta}} \gets \vec{\theta}$
\RETURN $\hat{\vec{\theta}}$
\end{algorithmic}
\end{algorithm}

\subsection{Fully Bayesian approach}\label{full_bayes}
In contrast to the empirical Bayes approach, one could fit a fully Bayesian model with the goal of finding the full posterior distribution of $\vec{\theta}\vert\mathbf{Y}$. Because there is no conjugate family for the hyperparameters $\vec{\theta}$, the posterior distribution is not available in a known closed form. Throughout, we use the diffuse normal priors from Sect.~\ref{sec:priors} on $\vec{\theta}$, which means that our unnormalized posterior is the integrated likelihood in Eq.~\eqref{eq:likelihood} times a normal density for each hyperparameter. A common approach in Bayesian modeling is to sample from the unnormalized posterior density $p(\vec{\theta} \vert \mathbf{Y})$, for example using importance sampling, MCMC, or Hamiltonian Monte Carlo. Regardless of the sampling approach taken, we are left with samples from the full posterior distribution of $\vec{\theta} \vert \mathbf{Y}$. However, many of these approaches are computationally intensive, especially when working with large spatial fields.
 
\subsection{Laplace approximation approach} \label{sec:laplace}
Because of the drawbacks of the previous two approaches (an empirical Bayes model does not account for any uncertainty in the map construction, while a fully Bayesian model is computationally prohibitive), we turn to the Laplace approximation. The Laplace approximation is a widely used technique for approximating the posterior distribution of a set of parameters. It involves approximating the posterior distribution of the hyperparameters $\vec{\theta}$ with a Gaussian distribution centered at the mode of the posterior:
\begin{equation}
    p(\vec{\theta} \mid \mathbf{Y}) \approx \hat{p}(\vec{\theta} \mid \mathbf{Y}) = \mathcal{N} \left(\hat{\vec{\theta}}, \mathbf{H}^{-1} \right)
\end{equation}
where $\hat{\vec{\theta}}$ is the maximum \emph{a posteriori} (MAP) estimate of $\vec{\theta}$, and $\mathbf{H}$ is the Hessian matrix of the negative log-posterior evaluated at $\hat{\vec{\theta}}$.
Using the same notation as in Eq.~\eqref{eq:L_tilde} for the negative log-posterior up to an additive constant, we have
\begin{equation}
    \hat{\vec{\theta}} = \underset{\vec{\theta}}{\operatorname{argmax}} \; p(\vec{\theta} \mid \mathbf{Y})  = \underset{\vec{\theta}}{\operatorname{argmin}} \, \tilde{\mathcal{L}}(\vec{\theta}),
    \qquad
    \mathbf{H} = \left[\nabla^2 \tilde{\mathcal{L}}(\vec{\theta})\right]_{\vec{\theta} = \hat{\vec{\theta}}}
    = \left[\nabla^2 \mathcal{L}(\vec{\theta})\right]_{\vec{\theta} = \hat{\vec{\theta}}} + \tau^{-2} \mathbf{I}.
    \label{eq:map_hessian}
\end{equation}

Finding $\hat{\vec{\theta}}$ here is identical to the estimation in the empirical Bayes approach in Sect.~\ref{emp_bayes}. By adding one relatively cheap operation at the end of this procedure, the Laplace approximation allows us to also estimate the uncertainty of the parameter estimates at a much smaller computational cost than the fully Bayesian approach. 
Specifically, once $\hat{\vec{\theta}}$ has been found, we calculate the Hessian matrix of the negative log-posterior evaluated at $\hat{\vec{\theta}}$, and then use this as the precision matrix of the approximate posterior distribution. For Bayesian transport maps, the Hessian matrix is not easily available in closed form, so we compute it using autodiff. The computational cost of this estimation does grow with $N$ and $n$; however, because we only require one such estimation and evaluation, the overall computational burden is only negligibly larger than that of the empirical Bayes approach. The steps for the Laplace approximation of the posterior are laid out in Algorithm~\ref{alg:laplace}.

\begin{algorithm}[htb]
\caption{Laplace approximation model fitting}
\label{alg:laplace}
\begin{algorithmic}[1]
\REQUIRE Training data $\mathbf{Y}$, prior $p(\vec{\theta})$, learning rate $\eta$, number of epochs $E$
\ENSURE Laplace approximation of the posterior of $\vec{\theta}$ given $\mathbf{Y}$: $\hat{p}(\vec{\theta} | \mathbf{Y}) = \mathcal{N}(\hat{\vec{\theta}}, \mathbf{H}^{-1})$
\STATE Initialize hyperparameters $\vec{\theta}$
\FOR{$e=1,2,\ldots,E$}
 \STATE Compute $\nabla_{\vec{\theta}} \tilde{\mathcal{L}}(\vec{\theta})$ using autodiff, where $\tilde{\mathcal{L}}(\vec{\theta})$ is defined in Eq.~\eqref{eq:L_tilde}
 \STATE $\vec{\theta} \gets \vec{\theta} - \eta \,\nabla_{\vec{\theta}} \tilde{\mathcal{L}}(\vec{\theta})$
\ENDFOR
\STATE $\hat{\vec{\theta}} \gets \vec{\theta}$
\STATE Compute  $\mathbf{H} = \big[\nabla^2 \tilde{\mathcal{L}}(\vec{\theta}) \big]_{\vec{\theta} = \hat{\vec{\theta}}}$ using autodiff
\RETURN $\hat{p}(\vec{\theta} \mid \mathbf{Y}) = \mathcal{N}(\hat{\vec{\theta}}, \mathbf{H}^{-1})$
\end{algorithmic}
\end{algorithm}

Formally, this approximation is derived from a second-order Taylor series expansion of the log-posterior density, $\log p(\vec{\theta}|\mathbf{Y})$, around the MAP estimate $\hat{\vec{\theta}}$. Defining $h(\vec{\theta}) = \log p(\vec{\theta}|\mathbf{Y})$, we have: 
\begin{equation}
    h(\vec{\theta}) \approx h(\hat{\vec{\theta}}) + \left(\vec{\theta} - \hat{\vec{\theta}} \right)^\top \nabla h(\hat{\vec{\theta}}) + \frac{1}{2} \left(\vec{\theta} - \hat{\vec{\theta}} \right)^\top \nabla^2 h(\hat{\vec{\theta}}) \left(\vec{\theta} - \hat{\vec{\theta}}\right)
\end{equation}
Since $\hat{\vec{\theta}}$ is the mode, the gradient $\nabla h(\hat{\vec{\theta}})$ is zero, and the first term $h(\hat{\vec{\theta}})$ is constant with respect to $\vec{\theta}$. Thus, exponentiating both sides reveals that the posterior is proportional to a Gaussian kernel with covariance matrix $\vec{\Sigma} = (-\nabla^2 h(\hat{\vec{\theta}}))^{-1}$, which corresponds to the inverse of the sum of the observed Fisher information matrix evaluated at the mode and the prior precision.

From a computational perspective, the validity of the approximation relies on the optimization procedure converging to a true optimum. Ending the optimization early by capping the number of epochs or stopping early based on validation metrics can leave a non-zero gradient $\nabla h(\hat{\vec{\theta}})$. If this term does not vanish in the Taylor expansion, the underlying assumption of the Laplace approximation is violated, which can lead to a poorly centered posterior with a misspecified covariance. Furthermore, the reliance on $\nabla h(\hat{\vec{\theta}})$ being zero presents challenges when working with stochastic optimization methods. While computing gradients on minibatches is standard practice when scaling to large datasets, the variance in these stochastic gradients may prevent the optimizer from achieving the mode, instead oscillating around it, leaving the final parameter estimate $\hat{\vec{\theta}}$ in a region where the full-data gradient is non-zero. To avoid this and ensure that the approximation is valid, it is often necessary to transition to a full-batch optimizer in the final stages of training to drive $\nabla h(\hat{\vec{\theta}})$ to zero. 

An additional challenge is the memory required when calculating the Hessian for large spatial datasets, as automatic differentiation of the full objective requires storing the computational graph for all $N$ locations at once. We avoid this by exploiting the linearity of the Hessian operator. The negative log integrated likelihood in Eq.~\eqref{eq:intlik} is a sum across all $N$ spatial locations, and the negative log-prior does not depend on the data and contributes
$\nabla^2\bigl(-\log p(\vec{\theta})\bigr) = \tau^{-2}\mathbf{I}$. Splitting the $N$ locations into $K$ disjoint batches $\mathcal{B}_1, \ldots, \mathcal{B}_K$, the full-data Hessian in
Eq.~\eqref{eq:map_hessian} can be written as 
\begin{equation}
    \mathbf{H} = \left[\nabla^2\tilde{\mathcal{L}}(\vec{\theta}) \right]_{\vec{\theta} = \hat{\vec{\theta}}} 
    = \sum_{k=1}^K \left[\nabla^2 \sum_{i\in \mathcal{B}_k} \mathcal{L}_i(\vec{\theta}) \right]_{\vec{\theta} = \hat{\vec{\theta}}} + \tau^{-2}\mathbf{I}
    = \sum_{k=1}^K \mathbf{H}_k + \tau^{-2}\mathbf{I} .
    \label{eq:batched_hessian}
\end{equation}
Each $\mathbf{H}_k$ can be computed by automatic differentiation using only the subset of locations in $\mathcal{B}_k$, allowing us to reconstruct $\mathbf{H}$ without constructing the full computational graph of our large spatial dataset in memory. 

\subsection{Generative modeling via posterior predictive sampling}
A key advantage of the fully Bayesian and Laplace approaches over empirical Bayes is the ability to propagate parameter uncertainty into the generated spatial fields. The posterior predictive distribution for a new field $\vec{y}_{\text{new}}$ is given by:
\begin{equation}
    p(\vec{y}_{\text{new}} \mid \mathbf{Y}) = \int p(\vec{y}_{\text{new}} \mid \vec{\theta}) p(\vec{\theta}|\mathbf{Y})\, \mathrm{d}\vec{\theta}
    \label{eq:pred}
\end{equation}
Under the empirical Bayes assumption, this integral collapses to $p \left(\vec{y}_{\text{new}} \mid \hat{\vec{\theta}}\right)$, ignoring the uncertainty in $\vec{\theta}$. In our Laplace framework, we approximate this integral using Monte Carlo sampling. We first draw $S$ samples of the hyperparameters from the approximate posterior, $\vec{\theta}^{(s)} \sim \mathcal{N}(\hat{\vec{\theta}}, \mathbf{H}^{-1})$. For each $\vec{\theta}^{(s)}$, we construct the corresponding transport map $\mathcal{T}_{\vec{\theta}^{(s)}}$ and generate a realization of the field by mapping samples from the reference distribution through the inverse map: $\vec{y}_*^{(s)} = \mathcal{T}^{-1}_{\vec{\theta}^{(s)}}(\vec{z}^{(s)})$, where $\vec{z}^{(s)} \sim \mathcal{N}(\vec{0}, \mathbf{I})$. This ensemble $\{\vec{y}_*^{(1)}, \dots, \vec{y}_*^{(S)}\}$ is a sample from the approximate posterior predictive distribution, which accounts for both the intrinsic stochasticity of the field and the epistemic uncertainty in the model parameters.

%% file: 04_numerical_comparison.tex
\section{Numerical comparison}\label{sec:numerical}

In this section, we assess the efficacy of the Laplace approximation for transport-map hyperparameters. 
We first compare the approximation to standard empirical Bayes and to reference posteriors obtained via MCMC in a controlled simulation environment (Sect.~\ref{sec:simulation_study}). Obtaining a reliable MCMC reference was itself challenging due to weak identifiability of some posterior directions for small training ensembles and discontinuities in the posterior. We describe these difficulties alongside our results. 

We then demonstrate the practical utility of the framework by constructing a stochastic surrogate for a large-scale climate field (Sect.~\ref{sec:application_data}), showing that our approach enables efficient generative modeling and uncertainty quantification in high-dimensional settings.

\subsection{Simulation study}\label{sec:simulation_study}
To explore the differences among the empirical Bayes, fully Bayesian, and proposed Laplace approximation methods, we carry out a simulation study. We use the \texttt{NR} scenario from \citet{Katzfuss02042024}, where the data-generating process can be characterized by a transport map in which each component is the sum of a linear and a nonlinear function. We let $\boldsymbol{\Sigma}$ be the exponential covariance matrix with entries $\Sigma_{k,l} = \exp(-\lVert \vec{s}_k - \vec{s}_l \rVert / 0.3)$, and let $y_{i,(j)}$ denote the value at the $j$th nearest previously ordered location to location $i$. 
We generate
$y_i = f_i(\vec{y}_{1:i-1}) + d_i \varepsilon_i$, with $\varepsilon_i \stackrel{\mathrm{iid}}{\sim} \mathcal{N}(0,1)$ and
\begin{equation}
    f_i(\vec{y}_{1:i-1}) = \left(\sum_{j=1}^{\min(i-1,\, 30)} b_{i,j}\, y_{i,(j)}\right)
    + 2\sin\bigl(4\,(b_{i,1}\, y_{i,(1)} + b_{i,2}\, y_{i,(2)})\bigr),
    \label{eq:nr}
\end{equation}
where $d_i^2 = \operatorname{Var}(y_i \mid \vec{y}_{1:i-1})$ and $b_{i,j}$ is the coefficient of the $j$th nearest previously ordered location in $\operatorname{E}(y_i \mid \vec{y}_{1:i-1})$,  computed under
$\mathcal{N}_N(\vec{0}, \boldsymbol{\Sigma})$. So the linear part of $f_i$ is the Gaussian conditional mean truncated to the 30 nearest previously ordered locations, and the nonlinear part
depends only on the two nearest.

\begin{figure}[htb]
    \centering
    \includegraphics[width =.98\linewidth]{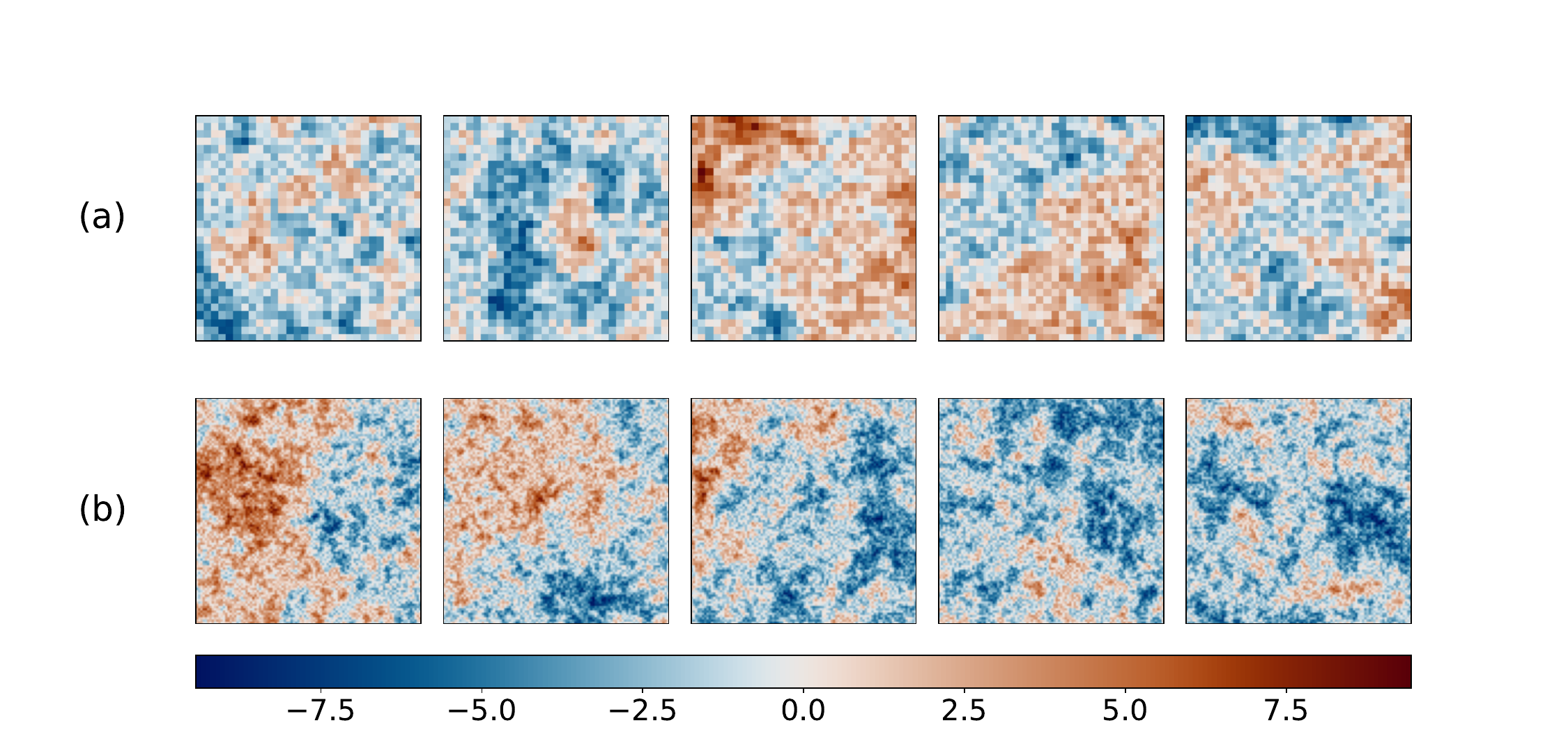}
    \caption{Selected simulated fields used in the simulation study, with (a) $N=900$ and (b) $N=10{,}000$ spatial locations.}
    \label{fig:sim_data}	
\end{figure}

We generated independent fields of size $N=900$ and $N=10{,}000$ spatial locations on a regular grid, and a selection of these simulated spatial fields is shown in Fig.~\ref{fig:sim_data}.

We considered settings where the transport map model was trained using $n=15, 20, 35, 50, 75, 100$ training samples, and evaluated performance on $20$ held-out test fields. Before fitting using all three methods (see Sect.~\ref{sec:methodology} and Algorithms~\ref{alg:empbayes} and~\ref{alg:laplace} for details), we centered both the training and test data location-wise using only the training fields, so that no information from the test fields enters the fitted model. Although the simulated fields have mean zero by construction, we center them in the same way as the real data in Sect.~\ref{sec:application_data}, so that both analyses follow the same procedure.

\subsubsection{Optimization and Laplace approximation} \label{sec:opt}
The posterior mode $\hat{\vec{\theta}}$ was found by minimizing the objective $\tilde{\mathcal{L}}(\vec{\theta})$ in three stages: a fast initial search using small batches, followed by a noise reduction step using larger batches to reduce stochastic variation, before ending with a full-batch polishing step. As described in Sect.~\ref{emp_bayes}, the objective is a sum over spatial locations, so minibatches of locations give unbiased estimates of the objective and its gradient. 

In the first step, we run 500 epochs of the Adam optimizer \citep{kingma2015adam} on minibatches of 128 and 1{,}024 locations for the $N=900$ and $N=10{,}000$ cases, respectively. With an initial learning rate $0.02$ halved twice at equal intervals, we move quickly from the default initial values toward the mode. The second stage consists of an additional 125 epochs of Adam with batches that are four times larger than in the previous stage. A learning rate of $2\times10^{-4}$ is used to reduce the stochastic noise. In the third and final stage, we run 30 epochs of L-BFGS \citep{Liu1989-mk} with a strong-Wolfe line search on the full data to locate the mode accurately.

After the third stage, the Hessian $\mathbf{H}$ of the negative log posterior at $\hat{\vec{\theta}}$ is computed by automatic differentiation, accumulated over batches of 100 locations as described in Eq.~\eqref{eq:batched_hessian}, to obtain the Laplace approximation $\mathcal{N}(\hat{\vec{\theta}}, \mathbf{H}^{-1})$. 

As a check on the optimization, we computed the Newton decrement $\lambda = \{\nabla\tilde{\mathcal{L}}(\hat{\vec{\theta}})^\top \mathbf{H}^{-1} \nabla\tilde{\mathcal{L}}(\hat{\vec{\theta}})\}^{1/2}$, which measures the remaining distance to the mode in units of the Laplace approximation's standard deviation. Across all fits reported here for the simulated data, $\lambda \leq 3.19 \times 10^{-5}$, meaning that the estimated distance to the mode is less than $10^{-4}$ posterior standard deviations for every parameter in $\vec{\theta}$.
The Hessian was positive definite at every reported estimate, which, together with the small $\lambda$ values, indicates that each estimate is a local mode and we were able to obtain a valid covariance matrix for the Laplace approximation.

\subsubsection{MCMC samplers} \label{sec:mcmc}
We compare the Laplace approximation to reference posteriors obtained by two adaptive Metropolis samplers \citep{Haario2001_new} that target the same posterior as the Laplace approximation. Both samplers have Gaussian random-walk proposals $\mathcal{N}(\vec{\theta}, \xi\, s_d\, \mathbf{C}_t)$, where $s_d = 2.38^2/d$
\citep{gelman1996efficient} for our $d=6$-dimensional parameter space, and the proposal covariance $\mathbf{C}_t$ is fixed at an initial value for the first $t_a$ iterations and then replaced by the empirical covariance of each chain's states after iteration $t_\ell$, plus a small diagonal regularization. The global scale $\xi$ is adapted by a Robbins--Monro update toward an acceptance rate of 0.234 \citep{Andrieu2008-vn, roberts1997weak}, so both adaptations diminish over time. 
The \emph{Hessian-free} sampler uses no output of the
optimization; chains start at or near the prior means with initial covariance $0.01\,\mathbf{I}$, and adaptation times $t_\ell = 2{,}000$ and $t_a = 4{,}000$. 
The \emph{Laplace-informed} sampler starts from overdispersed draws from $\mathcal{N}(\hat{\vec{\theta}}, 4\mathbf{H}^{-1})$ and uses the Laplace covariance $\mathbf{H}^{-1}$ as its initial proposal covariance, with shorter adaptation times $t_\ell = 500$ and $t_a = 1{,}000$. 

Each sampler runs four chains, with burn-in periods of 10{,}000 (Hessian-free) or 5{,}000 (Laplace-informed) iterations, followed by 10{,}000 saved iterations per chain. 
We assess convergence with the rank-normalized split-$\hat{R}$ and effective sample sizes (ESS) of
\citet{vehtari2021rank}, computed with ArviZ \citep{Kumar2019}. For the $N=10{,}000$ setting, the maximum $\hat{R}$ was below $1.01$ in every case except the Hessian-free sampler at $n = 100$ ($\hat{R} = 1.012$), and for all values of $n$, both the bulk and tail ESS were above 400. This suggests that both samplers performed well in the $N=10{,}000$ setting.

When using the $N=900$ data, both samplers converged for $n \geq 35$, with maximum $\hat{R}$ below $1.01$ and both bulk and tail ESS above 400. However, with small training sizes, obtaining trustworthy MCMC references proved difficult, and these difficulties are themselves informative about the posterior. First, the posterior has multiple weakly identified directions for small $n$. Even with a diffuse prior, for small $n$ the posterior extends along weakly identified directions. Chains started in different places explored different extensions, and mixing between them was slow. Near the transition to well-identified posteriors (which we take to be $n=15$), the main difficulty was a skewed marginal for $\theta_{d_2}$, which reduced the ESS to less than 100 and inflated the $\hat{R}$ values to $1.72$ for the Hessian-free sampler and $1.29$ for the Laplace-informed sampler. For these settings, we treat the MCMC output as a qualitative reference for the shape of the posterior rather than as an exact benchmark.

Second, there are discontinuities in the posterior that stem from the truncated conditioning sets $c(i)$. The number of active neighbors $m$ is the number of previously ordered neighbors $j \le m_{\max}$ whose relevance (see Sects.~\ref{subsec::ordering} and~\ref{sec:priors}) is at least $0.01$. This makes the likelihood piecewise smooth in $\theta_q$, with jumps where a neighbor enters or leaves. For the simulated domains, these jumps are small, but they grow with the number of locations, and they can create a separate local mode for each value of $m$, which random-walk samplers cannot leave. We return to this in Sect.~\ref{sec:application_data}.

\subsubsection{Pairwise posterior comparisons} \label{sec:pairpost}
With all three algorithms run, we start our analysis by examining the posterior distribution of $\vec{\theta}$ under all three models.

\begin{figure}[htb]
    \centering
    \includegraphics[width=0.83\textwidth]{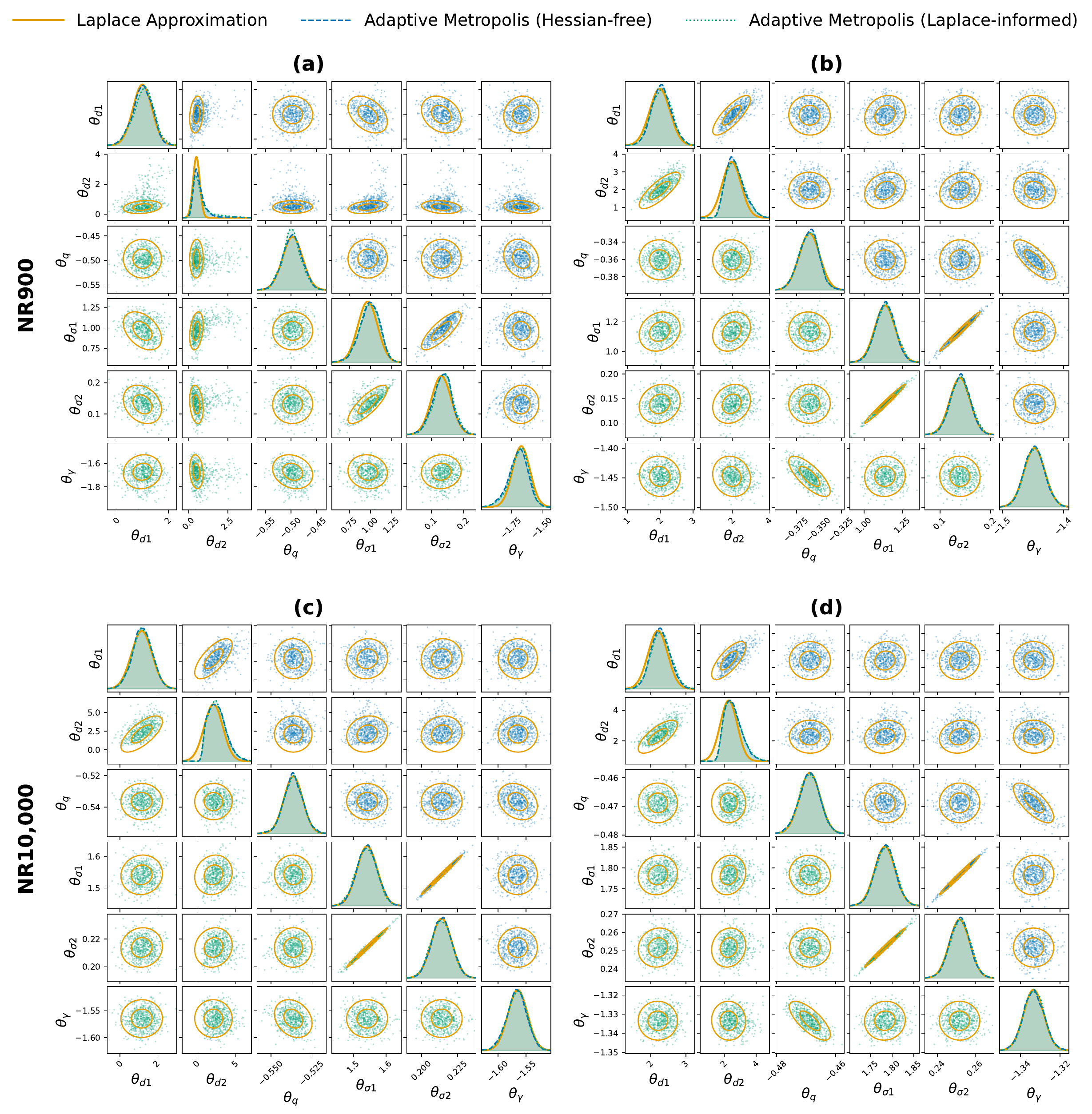} 
  \caption{Pairwise posterior distributions of $\vec{\theta}$ for (a) $N=900$, $n=20$; (b) $N=900$, $n=75$; (c) $N=10{,}000$, $n=20$; and (d) $N=10{,}000$, $n=75$. Note the comparison between the MCMC posteriors (blue and green) and the Laplace approximation Gaussian posterior (orange).}
  \label{fig:compare}
\end{figure}

First, we compare the joint posterior of $\vec{\theta}$ under each of the different approaches as $n$, the number of training samples, and $N$, the number of spatial locations, vary. To do this, we visualize the pairwise distributions of the hyperparameters in $\vec{\theta}$ under each regime. Figure~\ref{fig:compare} shows the pairwise posterior distributions of $\vec{\theta}$ for various scenarios and training ensemble sizes. Results for all settings are available in the publicly available code repository (see Code and data availability below). 

Figure~\ref{fig:compare} gives us valuable insight into how the posterior distribution changes as a function of $n$ and $N$. 
First, when $n$, the training ensemble size, is small, both the \emph{Hessian-free} and \emph{Laplace-informed} samplers show strong skew and non-Gaussianity in the marginal posterior of $\theta_{d_2}$. However, once the ensemble of training fields is large, the posterior distribution of $\vec{\theta}$ appears very Gaussian, as shown in Fig.~\ref{fig:compare}b and d, suggesting that the Laplace approximation is adequate.
We also found that the larger the spatial field, the fewer training samples are needed to learn the distribution, and the posterior appears more Gaussian (the marginal of $\theta_{d_2}$ is less skewed) with fewer samples, as shown in Fig.~\ref{fig:compare}c. So, when enough training data are available, the Laplace approximation appears to be an efficient way to accurately represent the posterior uncertainty in $\vec{\theta}$. However, as the amount of data used to estimate the map diminishes, the posterior looks less and less Gaussian, implying that the Laplace approximation may not be a good surrogate for the true posterior. 

\subsubsection{Map behavior: sparsity and nonlinearity}
While the direct posterior distributions of $\vec{\theta}$ are available from each method, and we examined them above, we note that these hyperparameters are difficult to interpret directly. So, we expand on our analysis by now focusing on the characteristics of the transport maps that are implied by the hyperparameters. Specifically, we will discuss how the posterior distribution on $\vec{\theta}$ implies a posterior distribution on map sparsity ($m$) and on the degree of nonlinearity in each $f_i$, controlled by $\sigma_i$.

\begin{figure}[!htbp]
	\centering
	\includegraphics[width=\textwidth]{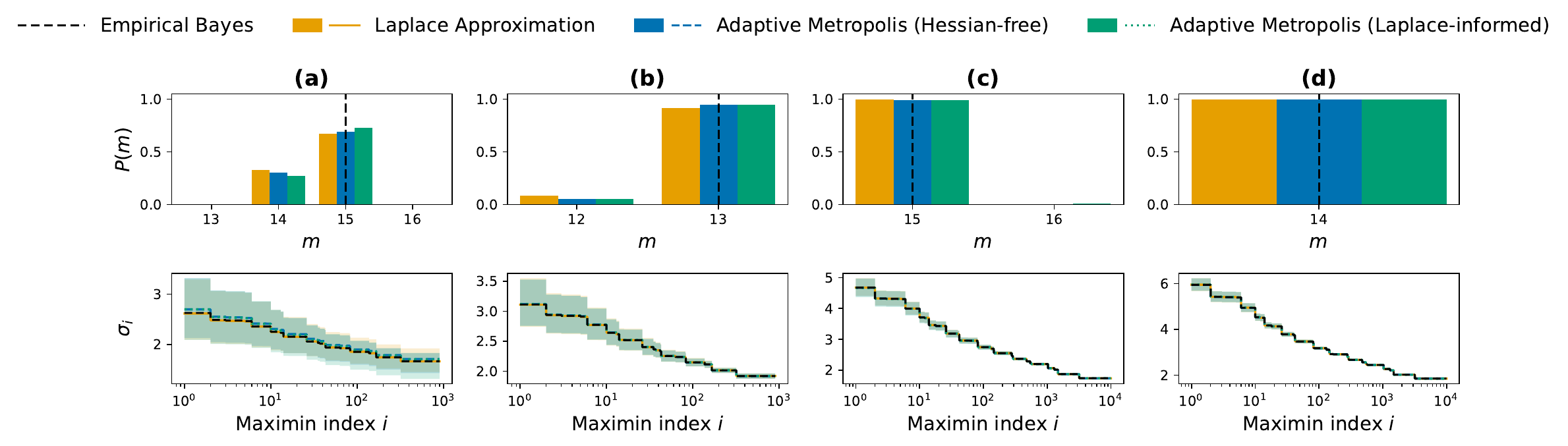}
	\caption{The first row shows the posterior distribution of the map sparsity $m$ under the Laplace approximation and MCMC. The second row shows the posterior medians and 95\,\% intervals of $\sigma_i$ under the Laplace approximation and MCMC, as a function of the index $i$ in the maximin ordering, for (a) $N=900$, $n=20$; (b) $N=900$, $n=75$; (c) $N=10{,}000$, $n=20$; and (d) $N=10{,}000$, $n=75$.}
	\label{fig:sparsity_nonlinearity}
\end{figure}

Figure~\ref{fig:sparsity_nonlinearity} shows the implied posterior distribution of $m$ for each method, as well as the posterior distributions of all the $\sigma_i$ under the same set of scenarios as in Fig.~\ref{fig:compare}. First, it is important to note that the posterior distribution of $m$ was almost identical under the fully Bayesian and Laplace approximation models, as shown in Fig.~\ref{fig:sparsity_nonlinearity}. Also, the final $m$ chosen by the empirical Bayes model was the mode of each of the full posterior distributions. Thus, we see that in terms of sparsity, the Laplace approximation is accurate, and that we are able to assess the variability in the level of sparsity in the transport map that was learned. 

Next, we examine the differences in the posterior distributions of $\sigma_i$. This represents the amount of nonlinearity that each $f_i$ has. Note that in all cases, $\sigma_i$ decreases along the maximin ordering, as we would expect. Second, we can see that there is much less uncertainty in each $\sigma_i$ when the amount of data available is large. We also see that when we have abundant training data, the distribution is very narrow, compared to a wider spread and larger variance in the posterior for each $\sigma_i$ when we have less data. We also see that the map appears to center around the same $\sigma_i$ values, and the largest change is in the spread of the distribution. So, it appears that the mean degree of nonlinearity is not affected by the amount of data available, while the variability around the degree of nonlinearity does change. We note that these plots show the posterior distribution of each $\sigma_i$ independently, and there may still be correlation between the $\sigma_i$.

\subsubsection{Log score comparison}
\label{subsec:logscores}
To compare the empirical Bayes, Laplace, and fully Bayesian approaches on held-out data, we evaluate the log of the posterior predictive density (see Eq.~\eqref{eq:pred}) of each method on a test set of $n_{\text{test}} = 20$ independent fields, $\vec{y}_{\text{test}}^{(1)}, \dots,
\vec{y}_{\text{test}}^{(n_{\text{test}})}$ generated from the same process as the training fields. 
As mentioned earlier, for the empirical Bayes method, the integral in the predictive distribution collapses to a plug-in distribution that depends only on $\hat{\vec{\theta}}$. 
\begin{equation}
    \mathrm{LS}_{\text{EB}}\!\left(\vec{y}_{\text{test}}^{(j)}\right)
    = \log p \left(\vec{y}_{\text{test}}^{(j)} | \hat{\vec{\theta}} \right).
    \label{eq:ls-eb}
\end{equation}

For the Laplace and Bayesian (MCMC) approaches, we approximate the integral in Eq.~\eqref{eq:pred} using a Monte Carlo sample of $S=10{,}000$ draws of $\vec{\theta}$ from the posterior. Therefore, these scores can be expressed as 
\begin{align}
    \mathrm{LS}\left(\vec{y}_{\text{test}}^{(j)}\right)
    = \log p\left(\vec{y}_{\text{test}}^{(j)} | \mathbf{Y}\right) 
    &= \log \left(\int p\left(\vec{y}_{\text{test}}^{(j)}|\vec{\theta}\right)p(\vec{\theta}|\mathbf{Y})\, \mathrm{d}\vec{\theta} \right) \nonumber \\
    &\approx \log \left(\frac{1}{S}\sum_{s=1}^{S} p\left(\vec{y}_{\text{test}}^{(j)} | \vec{\theta}^{(s)}\right)\right) \nonumber \\
    &= \operatorname{logsumexp}_{s=1}^{S}\left[\log p\left(\vec{y}_{\text{test}}^{(j)} | \vec{\theta}^{(s)}\right)\right] - \log S,
    \label{eq:ls-mc}
\end{align}
where $\operatorname{logsumexp}_{s} a_s := \log \sum_s \exp(a_s)$ helps with numerical stability.

For all methods, we summarize predictive performance by averaging the log scores over the $n_{\text{test}} = 20$ test fields that were generated using the same process as the training fields.

\begin{figure}[!htbp]
  \centering
  \includegraphics[width=\textwidth]{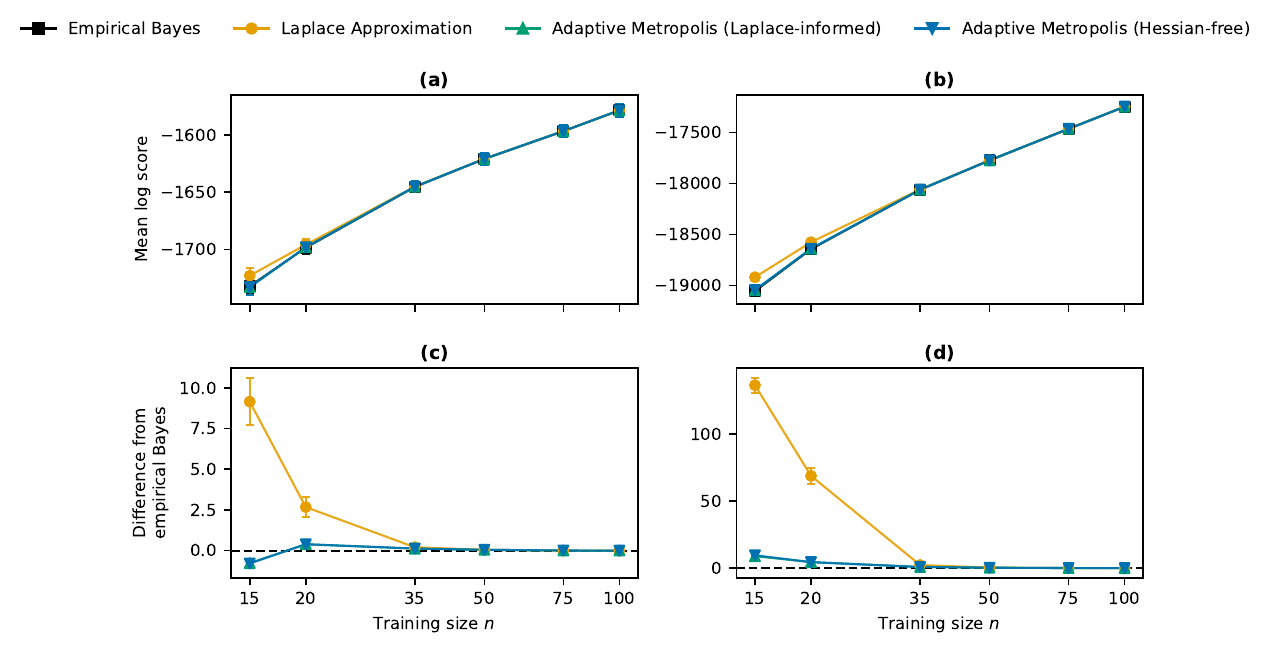}
  \caption{The top row shows the average log score as a function of the number of training fields $n$, for (a) $N=900$ and (b) $N=10{,}000$ with 95\,\% bootstrap intervals over the $n_{\text{test}} = 20$ test fields. Higher log scores are better. The bottom row shows the difference between each method and the standard empirical Bayes approach.}
  \label{fig:logscores}
\end{figure}

Figure~\ref{fig:logscores} compares the held-out scores for $n \geq 15$. With the exception of MCMC for $N=900$ with $n=15$ training samples, for which convergence was suspect as discussed in Sect.~\ref{sec:mcmc}, accounting for uncertainty in $\vec{\theta}$ was never noticeably worse than empirical Bayes; both the Laplace approximation and MCMC slightly improved the scores for small $n$. However, the scoring gains shrink quickly as $n$ increases, and by  $n = 35$, all methods from our simulation study score closely to each other. 
This is consistent with \citet{Katzfuss02042024}, who found little loss in inferential ability and accuracy when using a plug-in empirical Bayes estimate relative to a fully Bayesian approach, especially once the number of training samples $n$ grows large. However, for small $n$, there is more uncertainty in the estimate $\hat{\vec{\theta}}$, and capturing that uncertainty does improve model performance.

Among the tested uncertainty quantification methods, the Laplace approximation also scored higher than both MCMC samplers at small $n$. This was most visible for $N = 10{,}000$, where it improved on empirical Bayes by $\approx 135$ and $\approx 70$ per test field at $n = 15$ and $20$, compared with improvements of $\approx 10$ and $\approx 5$ for MCMC, even though the three posteriors are nearly indistinguishable in Fig.~\ref{fig:compare}c. 

Due to bias and Monte Carlo error in approximating these scores with autocorrelated Markov chains, we do not interpret the higher scores as evidence that the Laplace approximation is closer to the true posterior than MCMC. Instead, we suggest that the Laplace approximation is a robust and inexpensive basis for posterior predictive sampling and scoring that does not require the tuning and computational power of MCMC.

\subsubsection{Computational efficiency and run time}

\begin{figure}[!htbp]
  \centering
  \includegraphics[width=\textwidth]{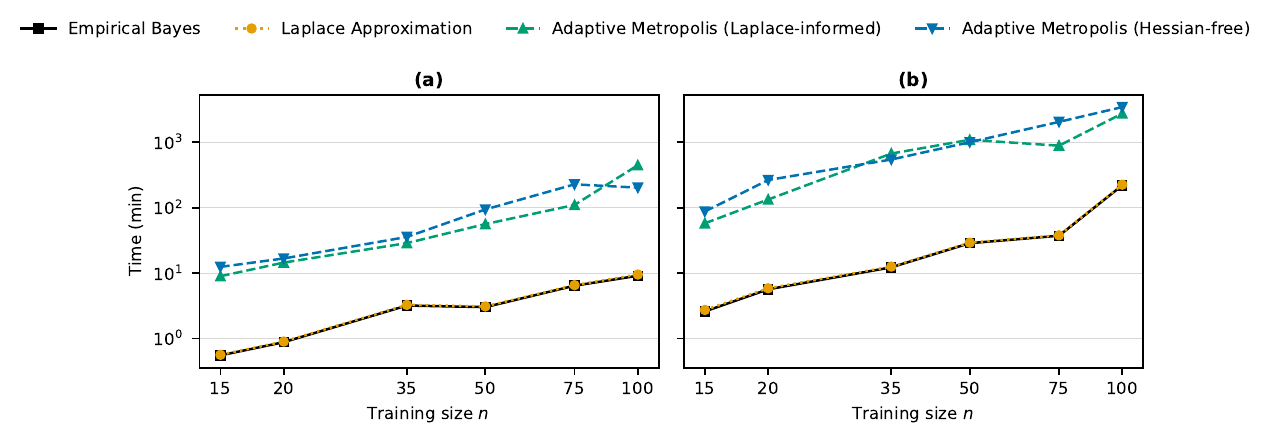}
  \caption{Wall-clock run time for each of the proposed algorithms. The reported empirical Bayes cost includes all three optimization stages, the Laplace cost adds the Hessian computation, and the MCMC methods report the cost to get $10{,}000$ post-burn-in samples.}
  \label{fig:time}
\end{figure}

The main advantage of the Laplace approximation over MCMC is its lower computational cost. While Fig.~\ref{fig:logscores} shows that the improvement in scoring performance decreases with $n$, each method achieves this performance at vastly different computational costs. Figure~\ref{fig:time} shows the wall-clock run times of each method as the training ensemble grows. 

\begin{figure}[!htbp]
  \centering
  \includegraphics[width=\textwidth]{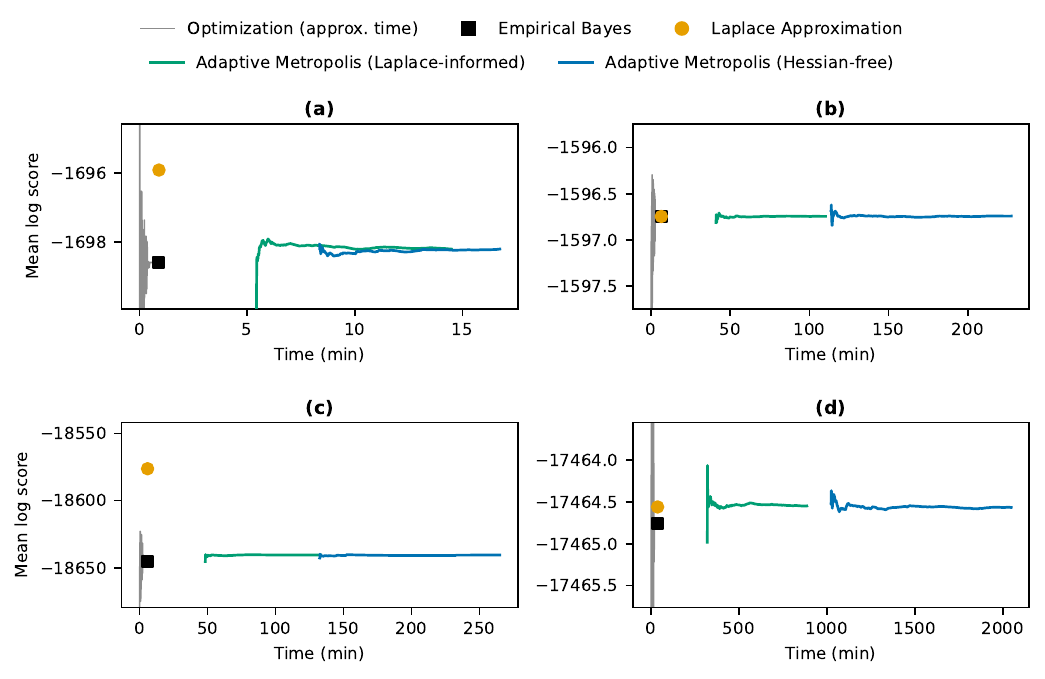}
  \caption{Mean held-out log score versus wall-clock time for (a) $N = 900$, $n = 20$; (b) $N = 900$, $n = 75$; (c) $N = 10{,}000$, $n = 20$; and (d) $N = 10{,}000$, $n = 75$. The gray line shows the plug-in log score at the current hyperparameter values during the three-stage optimization. The black square marks the final empirical Bayes estimate. The orange circle marks the Laplace approximation, whose time includes the Hessian computation. The green and blue lines show the log scores of the two adaptive Metropolis samplers, computed from the post-burn-in draws saved up to each time.}
  \label{fig:logscore_time}
\end{figure}

Seeing the vast difference in computation time, we were interested in combining each method's predictive performance with its computational cost. Figure~\ref{fig:logscore_time} does this by plotting each method's mean log score against the wall-clock time needed to obtain it. 

First, we note that the optimization reaches nearly its final plug-in score very quickly. Most of the iterations are used to drive the gradient to zero, which only changes the score slightly, but is necessary for the Laplace approximation to be valid. Second, as described in Sect.~\ref{sec:laplace}, the Hessian computation adds almost no time, so the Laplace approximation is available at essentially the same time as the empirical Bayes estimate, yet it scores better. Third, even the first MCMC draws after burn-in arrive several times later than the Laplace approximation. Finishing the samplers takes longer still, and their running scores remain flat at or below the score from the Laplace approximation, showing that the additional computation does not improve predictive performance.

Because the cost of the Laplace approximation relative to empirical Bayes is only the relatively fast Hessian computation, we find it to be an effective way to provide uncertainty quantification for the map's sparsity and nonlinearity, and improve predictive scoring, at a small fraction of the cost of MCMC.

\subsection{Real-data application}\label{sec:application_data}

\begin{figure}[!htbp]
  \centering
  \includegraphics[width=\textwidth]{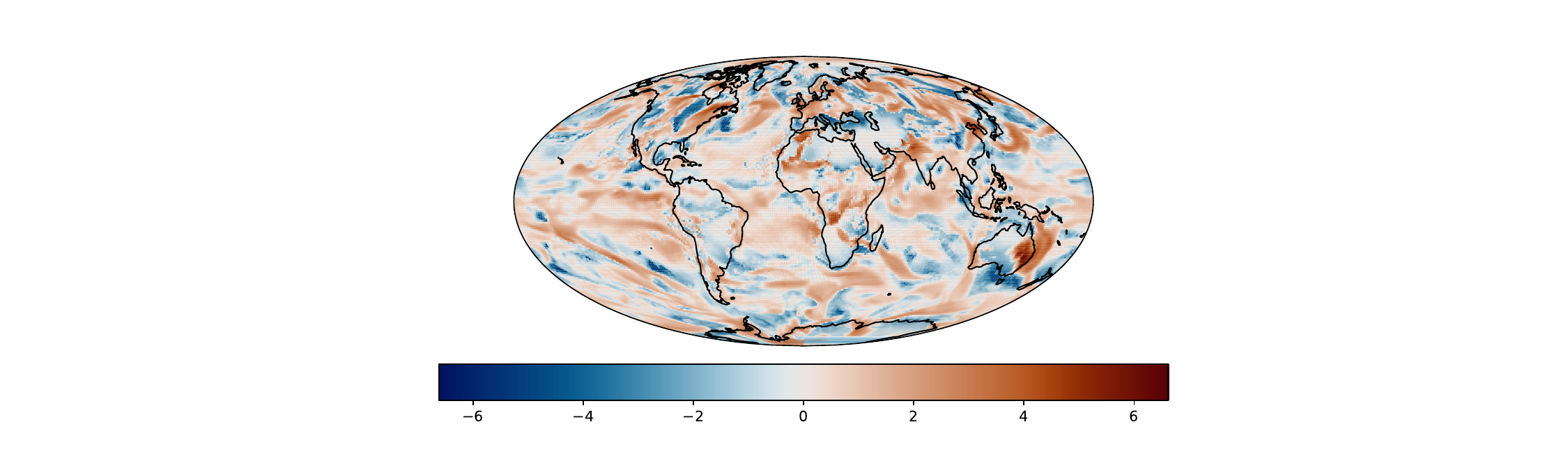}
  \caption{Example location-wise centered log-precipitation field consisting of $N=55{,}296$ locations across the globe.}
  \label{fig:precip}
\end{figure}

To demonstrate the utility of the Laplace approximation in an application with large spatial fields, we fit all three methods to log-precipitation rate fields from the Community Earth System Model (CESM) Large Ensemble Project \citep{Kay2015}. These fields represent one component of a larger model that simulates the full climate, but we restrict this work to the anomalies of the univariate log-precipitation rate field over $N=55{,}296$ locations on the globe in a $192 \times 288$ grid. Each field represents the 1 July daily-average field from consecutive CESM model years, and an example of one such field is shown in Fig.~\ref{fig:precip}. 

In a numerical experiment similar to the one described in Sect.~\ref{sec:simulation_study}, we fit the transport map with all three uncertainty quantification methods using $n=15$, 20, 35, and 50 training fields before evaluating performance on $n_{\text{test}}=20$ test fields. We did not consider training ensembles larger than $n=50$ due to the large computational cost of doing MCMC in this big data setting. Additionally, the \emph{Hessian-free} adaptive sampler struggled to converge (with bulk effective sample sizes less than 5 in some cases, even when using 10{,}000 post-burn-in draws), so we only use the \emph{Laplace-informed} MCMC samples in this section. For all four training sizes, the Laplace-informed chains had a maximum $\hat{R}$ of at most $1.01$ and minimum bulk and tail effective sample sizes of at least $1{,}166$ and $822$, respectively.

While the Laplace-informed MCMC chains appear to have converged for all training sizes, the optimizer did not converge equally well at each value of $n$. The Newton decrement was small at $n=20$ ($\lambda = 0.022$), but not at $n = 15$, 35, and 50 ($\lambda = 0.77$, 4.90, and 2.78). As discussed below, the log-posterior is discontinuous in $\theta_q$, as it can change significantly where a neighbor enters or leaves the conditioning sets used in the likelihood. With $n=35$ training fields, $\hat{\theta}_q$ lies at a discontinuity, and for $n=15$ and $n=50$ it lies within two posterior standard deviations of one.
If the optimum is at such a discontinuity, the gradient of the log-posterior in $\theta_q$ need not be zero, but the gradient should be zero in all other parameters. So, we also computed the Newton decrement over the other five hyperparameters with $\theta_q$ held fixed, but it remained nearly unchanged ($\lambda = 0.76$, 0.022,
3.95, and 2.78 for $n=15$, 20, 35, and 50), implying that the remaining gradient lies mostly in the smooth directions and that the optimization did not fully converge. We believe that this could be due to the final L-BFGS stage failing when an optimization step would cross a discontinuity, stopping progress in all directions and leaving the Laplace approximation distribution centered slightly away from the true mode.

\begin{figure}[!htbp]
  \centering
  \includegraphics[width=\textwidth]{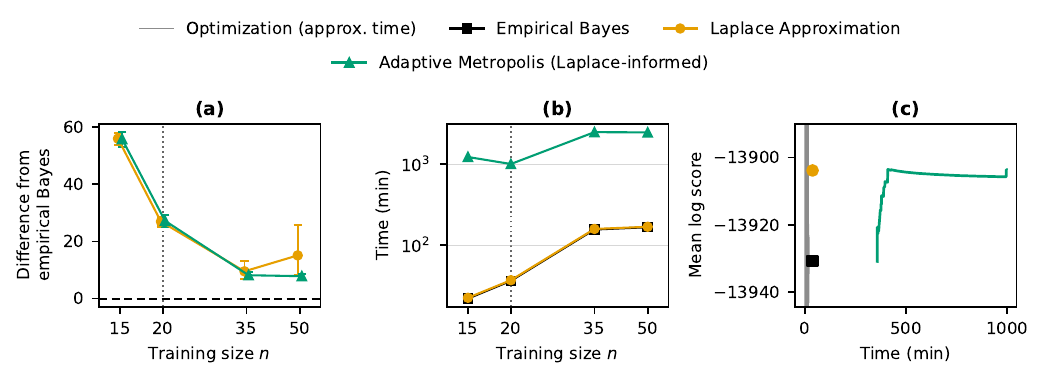}
  \caption{Scoring and timing for the log-precipitation application ($N = 55{,}296$). (a) Mean difference in held-out log score from empirical Bayes, with 95\,\% bootstrap intervals over the $n_{\text{test}} = 20$ test fields; (b) wall-clock time as a function of the number of training fields $n$. (c) Mean held-out log score versus wall-clock time for $n = 20$ (dotted line in (a) and (b)). In (c), the gray line shows the plug-in log score during the optimization; the square and circle mark empirical Bayes and the Laplace approximation, and the green line shows the cumulative log score over the post-burn-in draws from the Laplace-informed adaptive Metropolis sampler. Times for the sampler include the optimization and Hessian computation it starts from, and the $y$ axis of (c) is restricted to the range of the final scores. Higher log scores are better.}
  \label{fig:precip_scores}
\end{figure}

The Laplace-informed sampler uses the Laplace approximation only for its starting values (which are overdispersed draws from $\mathcal{N}(\hat{\vec{\theta}}, 4\mathbf{H}^{-1})$) and its initial proposal covariance. After the adaptation period, the proposal covariance is estimated from the chains themselves, so a mis-centered Laplace approximation may slow convergence, but should not change the distribution being sampled. 
Additionally, these offsets should have little effect on the behavior of the fitted maps themselves. Near the optimum, $\lambda^2/2$ approximates the change in the negative log-posterior that a Newton step would achieve. With $\theta_q$ held constant to account for the discontinuity, this is at most 7.8 (for $n=35$), or roughly $0.23$ per training field, which is small when compared to the magnitude of the training log-posterior ($1.2\times10^{6}$) and the differences in held-out log scores between the empirical Bayes and other methods, where the Laplace approximation and MCMC perform similarly (see Fig.~\ref{fig:precip_scores}). Due to the small nature of this mismatch and the similarity in the resulting held-out scores (see Fig.~\ref{fig:precip_scores}), we report the results as obtained via the same three-stage procedure described in Sect.~\ref{sec:opt}. Suggestions of how this could be resolved are detailed in Sect.~\ref{sec:conclusion}.

\begin{figure}[!htbp]
  \centering
  \includegraphics[width=\textwidth]{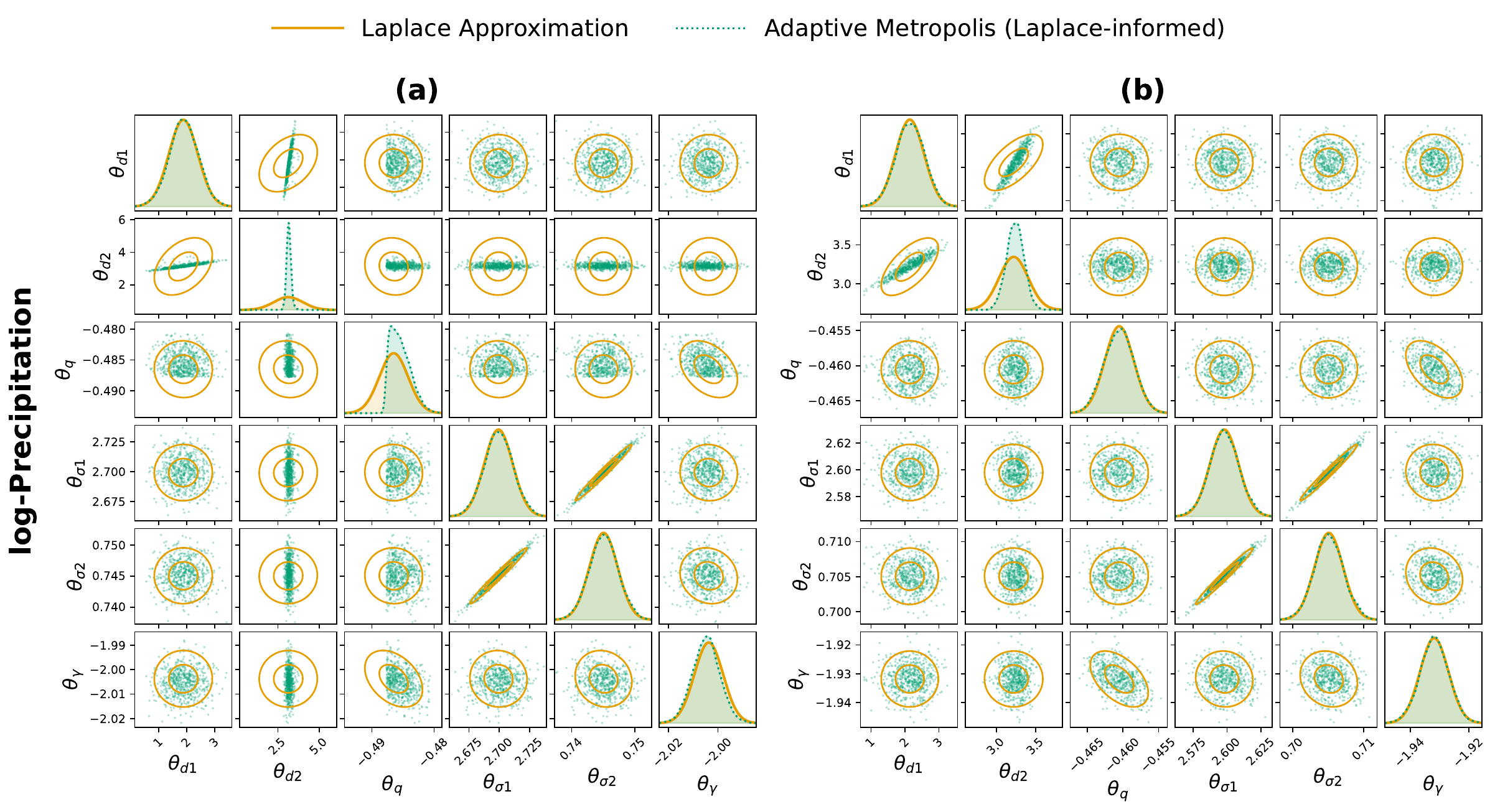}
  \caption{Pairwise posterior distributions of $\vec{\theta}$ for the application data with (a) $n=15$ and (b) $n=20$.}
  \label{fig:precip_pairs}
\end{figure}

Figure~\ref{fig:precip_pairs} shows the pairwise posteriors for $n=15$ and $n=20$, the two settings with the lowest Newton decrement. The $n=20$ setting shows that where the optimizer reached an interior mode far from the $\theta_q$ cutoffs, the Laplace approximation marginals agree closely with the MCMC posterior of all parameters except in the case of $\theta_{d_1}$ and $\theta_{d_2}$, where it is overdispersed. The $n=15$ setting in Fig.~\ref{fig:precip_pairs}a shows this same discrepancy in the joint distribution of $\theta_{d_1}$ and $\theta_{d_2}$, while also misrepresenting the marginal variance of $\theta_{d_2}$. 

These discrepancies, along with the strong marginal skew for $\theta_{d_2}$ discussed in Sect.~\ref{sec:pairpost}, suggest that a reparameterization of $\theta_{d_2}$ may be required for more accurate uncertainty quantification. Using $\exp(\theta_{d_2})$ as the coefficient of $\log\ell_i$ in Eq.~\eqref{eq:prior_di} instead of $\theta_{d_2}$ may help correct some of the approximation issues. 

In addition to misrepresenting the variance of $\theta_{d_2}$, when $n=15$, the Laplace approximation also misrepresents the marginal of $\theta_q$. This discrepancy is a consequence of how $m$, the size of the conditioning sets $c(i)$, is defined; $m = \left\lfloor -2 \log(\epsilon)\, e^{-\theta_q} \right\rfloor$ is an integer-valued step function of $\theta_q$. This means that the $\theta_q$ axis can be partitioned into intervals of constant $m$. Within each interval, the log-posterior is smooth, but there are jumps wherever $\theta_q$ crosses the threshold to add or remove a neighbor from $c(i)$. Because the log integrated likelihood sums over all $n$ training fields and $N$ locations, these jumps can be large, but the gradient-based optimizer and Laplace approximation are blind to the jumps because the Hessian is evaluated on the smooth part of the posterior at $\hat{\vec{\theta}}$. 
This is evident in the sharp marginal posterior for $\theta_q$ when sampled via MCMC, when compared to the Laplace approximation that puts substantial mass on either side of the cutoff between $m=14$ and $m=15$ at $\theta_q=-0.4877$.

\begin{figure}[!htbp]
  \centering
  \includegraphics[width=\textwidth]{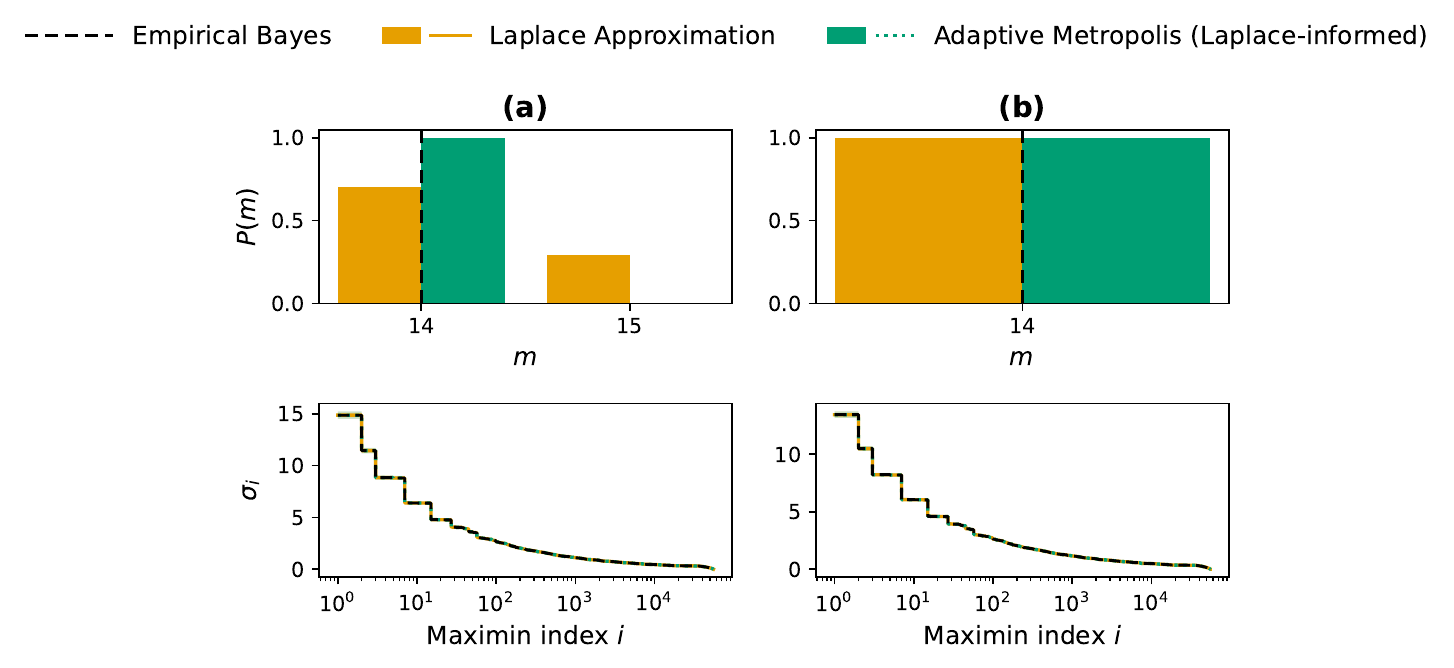}
  \caption{The top row shows the posterior distribution of the map sparsity $m$ under the Laplace approximation and MCMC for the application data with (a) $n=15$ and (b) $n=20$. The bottom row shows posterior medians and 95\,\% intervals of $\sigma_i$ under the Laplace approximation and MCMC, as a function of the index $i$ in the maximin ordering, for the application data under the same settings.}
  \label{fig:precip_m_sigma}
\end{figure}

Figure~\ref{fig:precip_m_sigma} shows the effect that this had on $m$. 
It is important to note that the \emph{Laplace-informed} MCMC chains started on both sides of this cutoff, but all ended up on the side corresponding to $m=14$ neighbors after burn-in, while the Laplace approximation splits its mass between the two conditioning set sizes. All other hyperparameters depend on the data only through $\ell_i$ and not through $m$, so their marginals, and functions computed from them such as $\sigma_i$, match much more closely between methods. 

Lastly, we explore the effect of the approximation on the log scores for the $n_{\text{test}}=20$ held-out fields, as shown in Fig.~\ref{fig:precip_scores}. At $n=20$, where we are confident that the model converged, we see that the Laplace approximation and MCMC methods both yield higher scores than plug-in empirical Bayes. For $n=15$, 35, and 50, where the Newton decrement is large and we did not reach a smooth interior mode (either because the mode lies at a discontinuity or because the optimizer stopped before reaching the mode), the empirical Bayes and Laplace scores may reflect the incomplete optimization. They do, however, still indicate that any optimization error has little effect on prediction, as the scores of both methods continue the trend from the simulated data setting and the Laplace approximation scores match the scores from MCMC. 
Good predictive performance does not imply that $\hat{\vec{\theta}}$ is at the mode or that the Laplace approximation describes the posterior of $\vec{\theta}$ accurately, but because the predictive scores are comparable to those of MCMC, it appears that the predictive distribution is quite insensitive to these parameter discrepancies. Once again, for prediction, the Laplace approximation remains a useful substitute for MCMC in this application, at a fraction of the cost.

%% file: 05_conclusion.tex
\conclusions\label{sec:conclusion}

In conclusion, we propose a practical, uncertainty-aware framework for the generative modeling of large non-Gaussian climate fields using triangular Bayesian transport maps. Instead of fixing the transport map parameters at point estimates obtained by optimization, we demonstrate that the distribution of the parameter vector $\vec{\theta}$ can be effectively approximated by a Laplace approximation, $\mathcal{N}(\hat{\vec{\theta}}, \mathbf{H}^{-1})$. This approach preserves the computationally efficient structure of the transport map while providing uncertainty quantification that is much cheaper than full MCMC, yet more informative than standard empirical Bayes approaches.

In our simulation study, we found that for moderate-to-large training sample sizes typical in climate science, the hyperparameter posterior is approximately Gaussian. Consequently, the Laplace approximation closely matches many aspects of reference MCMC posteriors and produces similar implied distributions on map sparsity and component nonlinearity. Furthermore, our real-data application to log-precipitation rates showed similar results for all hyperparameters except $\theta_q$, which has a discontinuous posterior, and confirmed that this framework scales effectively to real-world climate datasets with complex non-Gaussian features. In both cases, the predictive scores of the Laplace approximation were comparable to or higher than those of the fully Bayesian posterior fitted via MCMC. The fast computation of the Laplace approximation when compared to MCMC demonstrates a favorable trade-off between computation and rigor.

While promising, the work in this paper does have some limitations. The Laplace approximation assumes near-normality of the posterior around $\hat{\vec{\theta}}$, so when the true posterior is multimodal, strongly skewed, heavy-tailed, or discontinuous, the approximation can misrepresent the uncertainty we set out to capture. Additionally, because the degree of non-Gaussianity depends on $n$, $N$, and the data itself, we can only make broad generalizations about when the Laplace approximation is appropriate. However, we did find that the key map characteristics, such as sparsity and nonlinearity, are similar, regardless of how much data is available, and that even when working with a discontinuous density in $\theta_q$, the Laplace approximation had good predictive scoring.
Accurate Hessian estimation can also be numerically sensitive, even for a low-dimensional $\vec{\theta}$, and implementation details (optimizer, step sizes, and finite precision) influence the estimate of $\hat{\vec{\theta}}$. If the optimization stops early or gets stuck at a local minimum, the Laplace approximation can be centered away from the mode. We observed this for the log-precipitation data, where the truncation of the conditioning sets makes the posterior discontinuous in $\theta_q$ and the final optimization stage did not fully converge. The Newton decrement, computed with $\theta_q$ held constant, proved a useful diagnostic for detecting such incomplete convergence, and helped show why the effect of the mis-centering was small on the fitted maps and held-out scores. 

Following from these limitations, there is room for further work in this area to develop additional diagnostics that flag when the proposed Laplace approximation is suspected to be invalid while accounting for the known issues with the transport map hyperparameter posterior, such as the discontinuity in $\theta_q$. For this discontinuity in $\theta_q$, further work could explore an additional optimization stage over the other five parameters with $\theta_q$ held fixed.
There is also room for future work to extend the Laplace approximation and this framework to other applications beyond spatial fields. These other situations may have a larger hyperparameter vector, which may present additional challenges to creating an uncertainty-aware model. Lastly, future work could be done to explore a hybrid approach in which the non-Gaussian parts of the hyperparameter vector are estimated using other approximation techniques that allow for non-Gaussianity. 

Other future work is particularly relevant to the growing intersection of statistical climatology and machine learning (ML). Beyond refining the optimization procedures, there is significant potential to extend this framework to multivariate \citep{Wiemann2023} climate applications (e.g., jointly modeling temperature and precipitation) or spatiotemporal generative models. Additionally, a promising direction would be to integrate our uncertainty-aware transport maps into hybrid physics-ML Earth system models, where they could serve as stochastic sub-grid-scale parameterizations or emulators that fully account for parameter uncertainty.

%% file: laplace_transport_map_arxiv.bbl
\begin{thebibliography}{33}
\providecommand{\natexlab}[1]{#1}
\providecommand{\url}[1]{\texttt{#1}}
\providecommand{\urlprefix}{}
\expandafter\ifx\csname urlstyle\endcsname\relax
  \providecommand{\doi}[1]{https://doi.org/\discretionary{}{}{}#1}\else
  \providecommand{\doi}{https://doi.org/\discretionary{}{}{}\begingroup
  \urlstyle{rm}\Url}\fi

\bibitem[{Amaral~Turkman et~al.(2019)Amaral~Turkman, Paulino, and
  Müller}]{Amaral_Turkman_Paulino_Müller_2019}
Amaral~Turkman, M.~A., Paulino, C.~D., and Müller, P.: Computational Bayesian
  Statistics: An Introduction, Institute of Mathematical Statistics Textbooks,
  Cambridge University Press, Cambridge, \doi{10.1017/9781108646185}, 2019.

\bibitem[{Andrieu and Thoms(2008)}]{Andrieu2008-vn}
Andrieu, C. and Thoms, J.: A tutorial on adaptive MCMC, Statistics and
  Computing, 18, 343--373, \doi{10.1007/s11222-008-9110-y}, 2008.

\bibitem[{Berger et~al.(2001)Berger, de~Oliveira, and
  Sans{\'{o}}}]{Berger2001_new}
Berger, J.~O., de~Oliveira, V., and Sans{\'{o}}, B.: {Objective Bayesian
  analysis of spatial correlated data}, Journal of the American Statistical
  Association, 96, 1361--1374, \doi{10.1198/016214501753382282}, 2001.

\bibitem[{Bevilacqua et~al.(2020)Bevilacqua, Caamaño-Carrillo, and
  Gaetan}]{https://doi.org/10.1002/env.2632}
Bevilacqua, M., Caamaño-Carrillo, C., and Gaetan, C.: On modeling positive
  continuous data with spatiotemporal dependence, Environmetrics, 31, e2632,
  \doi{10.1002/env.2632}, 2020.

\bibitem[{Carlier et~al.(2010)Carlier, Galichon, and
  Santambrogio}]{doi:10.1137/080740647}
Carlier, G., Galichon, A., and Santambrogio, F.: From Knothe's Transport to
  Brenier's Map and a Continuation Method for Optimal Transport, SIAM Journal
  on Mathematical Analysis, 41, 2554--2576, \doi{10.1137/080740647}, 2010.

\bibitem[{Castruccio et~al.(2014)Castruccio, McInerney, Stein, Crouch, Jacob,
  and Moyer}]{Castruccio2014}
Castruccio, S., McInerney, D.~J., Stein, M.~L., Crouch, F.~L., Jacob, R.~L.,
  and Moyer, E.~J.: {Statistical emulation of climate model projections based
  on precomputed GCM runs}, Journal of Climate, 27, 1829--1844,
  \doi{10.1175/JCLI-D-13-00099.1}, 2014.

\bibitem[{{Center for High Throughput Computing}(2006)}]{CHTC2006}
{Center for High Throughput Computing}: Center for High Throughput Computing,
  \doi{10.21231/GNT1-HW21}, 2006.

\bibitem[{Chakraborty and
  Katzfuss(2026)}]{chakraborty2024learningnongaussianspatialdistributions}
Chakraborty, A. and Katzfuss, M.: Learning non-{G}aussian spatial distributions
  via {B}ayesian transport maps with parametric shrinkage, Journal of
  Agricultural, Biological and Environmental Statistics, 31, 641--659,
  \doi{10.1007/s13253-025-00687-5}, 2026.

\bibitem[{Eskelson et~al.(2011)Eskelson, Madsen, Hagar, and
  Temesgen}]{10.1093/forestscience/57.3.212}
Eskelson, B. N.~I., Madsen, L., Hagar, J.~C., and Temesgen, H.: Estimating
  Riparian Understory Vegetation Cover with Beta Regression and Copula Models,
  Forest Science, 57, 212--221, \doi{10.1093/forestscience/57.3.212}, 2011.

\bibitem[{Fonseca et~al.(2023)Fonseca, Lobo, and
  Schmidt}]{10.1093/jrsssc/qlac007}
Fonseca, T. C.~O., Lobo, V. G.~R., and Schmidt, A.~M.: Dynamical non-Gaussian
  modelling of spatial processes, Journal of the Royal Statistical Society
  Series C: Applied Statistics, 72, 76--103, \doi{10.1093/jrsssc/qlac007},
  2023.

\bibitem[{Gelman et~al.(1996)Gelman, Roberts, and Gilks}]{gelman1996efficient}
Gelman, A., Roberts, G.~O., and Gilks, W.~R.: Efficient Metropolis Jumping
  Rules, in: Bayesian Statistics 5: Proceedings of the Fifth Valencia
  International Meeting, edited by Bernardo, J.~M., Berger, J.~O., Dawid,
  A.~P., and Smith, A. F.~M., Oxford University Press, ISBN 9780198523567,
  \doi{10.1093/oso/9780198523567.003.0038}, 1996.

\bibitem[{Guinness(2018)}]{Guinness2016a}
Guinness, J.: {Permutation and grouping methods for sharpening Gaussian process
  approximations}, Technometrics, 60, 415--429,
  \doi{10.1080/00401706.2018.1437476}, 2018.

\bibitem[{Haario et~al.(2001)Haario, Saksman, and Tamminen}]{Haario2001_new}
Haario, H., Saksman, E., and Tamminen, J.: An Adaptive Metropolis Algorithm,
  Bernoulli, 7, 223--242, \doi{10.2307/3318737}, 2001.

\bibitem[{Heaton et~al.(2019)Heaton, Datta, Finley, Furrer, Guinness,
  Guhaniyogi, Gerber, Gramacy, Hammerling, Katzfuss, Lindgren, Nychka, Sun, and
  Zammit-Mangion}]{Heaton2017}
Heaton, M.~J., Datta, A., Finley, A.~O., Furrer, R., Guinness, J., Guhaniyogi,
  R., Gerber, F., Gramacy, R.~B., Hammerling, D.~M., Katzfuss, M., Lindgren,
  F., Nychka, D.~W., Sun, F., and Zammit-Mangion, A.: {A case study competition
  among methods for analyzing large spatial data}, Journal of Agricultural,
  Biological, and Environmental Statistics, 24, 398--425,
  \doi{10.1007/s13253-018-00348-w}, 2019.

\bibitem[{Hobert and Casella(1996)}]{Hobert01121996}
Hobert, J.~P. and Casella, G.: The Effect of Improper Priors on Gibbs Sampling
  in Hierarchical Linear Mixed Models, Journal of the American Statistical
  Association, 91, 1461--1473, \doi{10.1080/01621459.1996.10476714}, 1996.

\bibitem[{Hörning and Bárdossy(2025)}]{HORNING2025100872}
Hörning, S. and Bárdossy, A.: Simulation of conditional non-Gaussian random
  fields with directional asymmetry, Spatial Statistics, 65, 100\,872,
  \doi{10.1016/j.spasta.2024.100872}, 2025.

\bibitem[{Johnson et~al.(2026)Johnson, Jimenez, and Katzfuss}]{Johnson2026code}
Johnson, J., Jimenez, F., and Katzfuss, M.: Code and data for ``{L}aplace
  {B}ayesian transport maps for generative modeling of non-{G}aussian climate
  fields'', Zenodo [code], \doi{10.5281/zenodo.24203962}, 2026.

\bibitem[{Katzfuss and Guinness(2021)}]{Katzfuss&Guinness2021}
Katzfuss, M. and Guinness, J.: A General Framework for Vecchia Approximations
  of Gaussian Processes, Statistical Science, 36, 124--141,
  \doi{10.1214/19-STS755}, 2021.

\bibitem[{Katzfuss and Schäfer(2024)}]{Katzfuss02042024}
Katzfuss, M. and Schäfer, F.: Scalable Bayesian Transport Maps for
  High-Dimensional Non-Gaussian Spatial Fields, Journal of the American
  Statistical Association, 119, 1409--1423,
  \doi{10.1080/01621459.2023.2197158}, 2024.

\bibitem[{Kay et~al.(2015)Kay, Deser, Phillips, Mai, Hannay, Strand, Arblaster,
  Bates, Danabasoglu, Edwards, Holland, Kushner, Lamarque, Lawrence, Lindsay,
  Middleton, Munoz, Neale, Oleson, Polvani, and Vertenstein}]{Kay2015}
Kay, J.~E., Deser, C., Phillips, A., Mai, A., Hannay, C., Strand, G.,
  Arblaster, J.~M., Bates, S.~C., Danabasoglu, G., Edwards, J., Holland, M.,
  Kushner, P., Lamarque, J.~F., Lawrence, D., Lindsay, K., Middleton, A.,
  Munoz, E., Neale, R., Oleson, K., Polvani, L., and Vertenstein, M.: {The
  Community Earth System Model (CESM) Large Ensemble Project: A community
  resource for studying climate change in the presence of internal climate
  variability}, Bulletin of the American Meteorological Society, 96,
  1333--1349, \doi{10.1175/BAMS-D-13-00255.1}, 2015.

\bibitem[{Kay et~al.(2021)Kay, Deser, Phillips, and
  Simpson}]{ncar_gdex_dataset_d651027}
Kay, J.~E., Deser, C., Phillips, A.~S., and Simpson, I.~R.: {CESM1 Large
  Ensemble Community Project}, NSF National Center for Atmospheric Research
  [data set], \doi{10.5065/D6J101D1}, 2021.

\bibitem[{Kingma and Ba(2015)}]{kingma2015adam}
Kingma, D.~P. and Ba, J.: Adam: A Method for Stochastic Optimization, in:
  International Conference on Learning Representations (ICLR),
  \urlprefix\url{https://arxiv.org/abs/1412.6980}, 2015.

\bibitem[{Kumar et~al.(2019)Kumar, Carroll, Hartikainen, and
  Martin}]{Kumar2019}
Kumar, R., Carroll, C., Hartikainen, A., and Martin, O.: ArviZ a unified
  library for exploratory analysis of Bayesian models in Python, Journal of
  Open Source Software, 4, 1143, \doi{10.21105/joss.01143}, 2019.

\bibitem[{Liu and Nocedal(1989)}]{Liu1989-mk}
Liu, D.~C. and Nocedal, J.: On the limited memory BFGS method for large scale
  optimization, Mathematical Programming, 45, 503--528,
  \doi{10.1007/BF01589116}, 1989.

\bibitem[{Marzouk et~al.(2016)Marzouk, Moselhy, Parno, and
  Spantini}]{Marzouk2016}
Marzouk, Y.~M., Moselhy, T., Parno, M., and Spantini, A.: {Sampling via measure
  transport: An introduction}, in: Handbook of Uncertainty Quantification,
  edited by Ghanem, R., Higdon, D., and Owhadi, H., Springer, ISBN
  9783319123851, \doi{10.1007/978-3-319-12385-1}, 2016.

\bibitem[{Nychka et~al.(2018)Nychka, Hammerling, Krock, and Wiens}]{Nychka2018}
Nychka, D.~W., Hammerling, D.~M., Krock, M., and Wiens, A.: {Modeling and
  emulation of nonstationary Gaussian fields}, Spatial Statistics, 28, 21--38,
  \doi{10.1016/j.spasta.2018.08.006}, 2018.

\bibitem[{Papamakarios et~al.(2021)Papamakarios, Nalisnick, Rezende, Mohamed,
  and Lakshminarayanan}]{Papamakarios2021}
Papamakarios, G., Nalisnick, E., Rezende, D.~J., Mohamed, S., and
  Lakshminarayanan, B.: Normalizing Flows for Probabilistic Modeling and
  Inference, Journal of Machine Learning Research, 22, 1--64,
  \urlprefix\url{http://jmlr.org/papers/v22/19-1028.html}, 2021.

\bibitem[{Paszke et~al.(2017)Paszke, Gross, Chintala, Chanan, Yang, DeVito,
  Lin, Desmaison, Antiga, and Lerer}]{Paszke2017AutomaticDI}
Paszke, A., Gross, S., Chintala, S., Chanan, G., Yang, E., DeVito, Z., Lin, Z.,
  Desmaison, A., Antiga, L., and Lerer, A.: Automatic differentiation in
  PyTorch, in: NIPS 2017 Workshop on Autodiff,
  \urlprefix\url{https://openreview.net/forum?id=BJJsrmfCZ}, 2017.

\bibitem[{Roberts et~al.(1997)Roberts, Gelman, and Gilks}]{roberts1997weak}
Roberts, G.~O., Gelman, A., and Gilks, W.~R.: Weak convergence and optimal
  scaling of random walk Metropolis algorithms, The annals of applied
  probability, 7, 110--120, \doi{10.1214/aoap/1034625254}, 1997.

\bibitem[{Sch{\"{a}}fer et~al.(2021)Sch{\"{a}}fer, Sullivan, and
  Owhadi}]{Schafer2017}
Sch{\"{a}}fer, F., Sullivan, T.~J., and Owhadi, H.: {Compression, inversion,
  and approximate PCA of dense kernel matrices at near-linear computational
  complexity}, Multiscale Modeling {\&} Simulation, 19, 688--730,
  \doi{10.1137/19M129526X}, 2021.

\bibitem[{Speagle(2020)}]{speagle2020conceptualintroductionmarkovchain}
Speagle, J.~S.: A Conceptual Introduction to Markov Chain Monte Carlo Methods,
  arXiv [preprint], \doi{10.48550/arXiv.1909.12313}, 2020.

\bibitem[{Vehtari et~al.(2021)Vehtari, Gelman, Simpson, Carpenter, and
  B{\"u}rkner}]{vehtari2021rank}
Vehtari, A., Gelman, A., Simpson, D., Carpenter, B., and B{\"u}rkner, P.-C.:
  {Rank-Normalization, Folding, and Localization: An Improved {$\widehat{R}$}
  for Assessing Convergence of MCMC (with Discussion)}, Bayesian Analysis, 16,
  667 -- 718, \doi{10.1214/20-BA1221}, 2021.

\bibitem[{Wiemann and Katzfuss(2023)}]{Wiemann2023}
Wiemann, P. F.~V. and Katzfuss, M.: Bayesian Nonparametric Generative Modeling
  of Large Multivariate Non-Gaussian Spatial Fields, Journal of Agricultural,
  Biological and Environmental Statistics, 28, 597--617,
  \doi{10.1007/s13253-023-00580-z}, 2023.

\end{thebibliography}
